\documentclass[reprint,amsmath,amssymb,prstab,a4paper]{revtex4-2}

\usepackage{float}
\makeatletter
\let\newfloat\newfloat@ltx
\makeatother

\usepackage{xcolor,graphicx,hyperref,siunitx,algorithm,algpseudocode}
\usepackage{cancel}
\usepackage{pgfplots}
\usepackage[caption = false]{subfig}
\usepackage[Export]{adjustbox}
\usepackage{booktabs}

\pgfplotsset{compat=newest}
\definecolor{uncomputed}{rgb}{0.651,0.792,0.941}
\algtext*{EndWhile}
\algtext*{EndIf}
\algtext*{EndFor}

\newcommand{\dpp}{(\mathrm d p /p)}
\newcommand{\Dppo}{(\Delta p /p_0)}
\newcommand\beq{\begin{equation}}
\newcommand\eeq{\end{equation}}

\newcommand{\powten}[1]{10^{#1}}

\begin{document}

\title{Efficient algorithms for dynamic aperture and momentum acceptance calculation}
\author{Bernard Riemann}
\author{Masamitsu Aiba}
\author{Jonas Kallestrup}
\email{Corresponding author\\ jonas.kallestrup@psi.ch}
\author{Andreas Streun}
\affiliation{Paul Scherrer Institut, CH-5232 Villigen PSI, Switzerland}

\begin{abstract}
New algorithms useful for the calculation of dynamic aperture and momentum acceptance in circular accelerators are developed and presented.
The flood-fill tool from raster graphics inspired us to efficiently compute dynamic apertures by minimizing required trackings on stable initial coordinates, leading to several factors of speed-up with respect to standard algorithms. A novel technique for momentum acceptance calculations, Fast Touschek Tracking, is developed. Thorough benchmarking using modern accelerator codes shows that the new technique can provide one or two orders of magnitude faster computation of local momentum acceptances with only limited loss of accuracy. 
\end{abstract}

\maketitle

\section{Introduction}
The next generation light sources (NGLS) achieve low natural emittance based on multi-bend achromat (MBA) lattices, thereby improving the photon beam performance drastically. Several facilities have already been constructed~\cite{MAXIV,ESRFEBS,Sirius}, while many more NGLS facilities are under commissioning, construction or planning~\cite{PhysRevAccelBeams.26.091601,APSU,HEPS,ALSU,PETRAIV,ELETTRA2,SOLEILU,SPRING82}.

Low natural emittance within a given circumference requires strong horizontal focusing to suppress dispersion in the bending magnets, i.e., at the locations of synchrotron radiation emission. Strong sextupoles have to be installed subsequently to correct the chromaticity, i.e., the chromatic aberrations of the quadrupole or combined function magnets used for focusing. The nonlinearity of the sextupole fields leads to limits of stable motion of stored electrons. Sextupoles are arranged in patterns with appropriate phase advances to cancel adverse nonlinear effects while maintaining the chromaticity correction, and more (harmonic) sextupoles and higher multipoles may be inserted to further suppress resonance driving terms (RDTs) and control both chromatic tunes shifts (CTS) and amplitude dependent tune shifts (ADTS). Chromatic tune shifts may move betatron tunes to a resonance at some momentum offset and thus limit the momentum acceptance, and the combination of CTS, ADTS and RDTs may lead to very small transverse on- and off-momentum acceptance. The optimization of these lattice properties by means of analytical and/or numerical approaches is a crucial step to design high-performance storage rings.

Furthermore, Touschek lifetime~\cite{PhysRevLett.10.407} is an important objective in the optimization of MBA storage ring lattices due to the high particle density of the low emittance beam. Touschek scattering describes a two-particle process, where two electrons transfer transverse momentum from betatron-oscillations to longitudinal momentum, such that one particle is accelerated and the other one is decelerated. In the lab system transverse momentum is small compared to longitudinal momentum, but this is not the case in the comoving system of the electron bunch, as a simple estimate may demonstrate: Transverse (horizontal) momentum typically is $p_x=p_0\sqrt{\varepsilon_x/\beta_x}$ with $p_0c=E_0$ the storage ring operational energy, $\varepsilon_x$ its horizontal emittance and $\beta_x$ the local beta-function. If $p_x$ turns completely to longitudinal momentum in the moving system, the resulting momentum offset in the lab system becomes $\Dppo=\gamma\sqrt{\varepsilon_x/\beta_x}$, with $\gamma$ the Lorentz factor. Typical parameters of middle-energy NGLS like $E_0=\SI{3}{\giga\electronvolt}$, $\varepsilon_x=\SI{100}{\pico\meter\radian}$ and $\beta_x=\SI{1}{\meter}$ give $\Dppo\approx 6\,{\rm \%}$, which presents a challenge to the optimization of off-momentum optics. 
After the scattering event not only the momenta of the particles have changed to $\pm\Dppo$, but the particles will also start betatron oscillations around the off-momentum closed orbits at $\pm\Dppo$, the oscillation amplitudes of which are determined by the optical functions at the location of the scattering event. Thus the machine's momentum acceptance (MA), i.e., the minimum/maximum energy deviation not leading to a particle loss, varies along the ring. Touschek lifetime is obtained from the local MA together with other relevant parameters. 

The dynamic aperture (DA), i.e., transverse acceptance, as required for injection and the local MA as required for Touschek lifetime are most prominent objectives in optimization but computationally expensive. A thorough and robust design using phase-cancellation schemes, suppression of low-order RDTs and linear ADTS typically yields decent DA and MA~\cite{bengtsson_sls,farvacque:ipac13-mopea008,bengtsson_sls2}, but further gains in performance can sometimes still be achieved by using numerical optimization algorithms based on multi-objective optimization or machine learning techniques (see, e.g., \cite{moga-ehrlichman,moga-li-cheng,mopso-huang,moga-kranjcevic,lu-ML} for a few of many examples).

These approaches numerically compute DAs and MA, aiming at improving them at the expense of computation time. It is noted that the computation of MA (and then Touschek lifetime) is time-consuming especially in the MBA lattice, where the optical functions are varying within short distances: The momentum acceptances need to be computed at many locations of the ring in order to accurately evaluate Touschek lifetime. Moreover, the computation of DA and MA is repeated many times (Monte Carlo simulation) to evaluate the influence of inevitable machine imperfections such as misalignments and magnetic field errors. Hence computationally efficient algorithms are required. 

To address these aforementioned computational issues, we present new algorithms that provide significant speed ups of DA and MA computation with a minimal loss of accuracy and precision.  

The paper is organized as follows: In Sec.~\ref{sec:DA} the computation of DA is discussed and the ``flood-fill'' algorithm is introduced. The performances of the various algorithms are quantitatively evaluated using real lattice examples. In Sec.~\ref{sec:TL} a new ``Fast Touschek Tracking'' method to compute the local momentum acceptance is described. Sec.~\ref{sec:results} highlights the implementation of
Fast Touschek Tracking in two common accelerator codes: OPA and Accelerator toolbox. Performance studies for lattices with realistic errors
and with full coupling are presented. Finally, we draw our conclusions and propose further developments in Sec.~\ref{sec:Conclusion} and Sec.~\ref{sec:Outlook}.

\section{Dynamic aperture}
\label{sec:DA}
The maximum stable betatron oscillation amplitude in a ring is limited by either the physical aperture or the dynamic aperture. The physical aperture is determined by physical objects such as vacuum chambers or collimators, while the dynamic aperture is limited by the nonlinear particle motion arising from mainly higher multipoles (sextupoles and octupoles). Strengths of the nonlinear magnets need to be selected carefully or optimized in order to make the DA sufficient. The DA is computed by tracking a particle for various initial conditions, i.e., scanning over the horizontal and vertical coordinates ($x$ and $y$) typically at a fixed longitudinal location of the ring. 

\subsection{Standard approaches}
The dynamic aperture is usually estimated via binary search (BS) that finds the boundary between stable and unstable initial coordinate on a line. It is repeated for several lines to visualize the DA in the two-dimensional plane, $x$--$y$. A set of radial lines (or rays) centered at the accelerator design axis or at the closed orbit is widely used. Such a computation is outlined in Alg.~\ref{alg:da_binary}.

\begin{algorithm}

\begin{algorithmic}
\For{every ray at a longitudinal location, $s_0$}
  \State $\vec X_0 \gets$ particle coordinates of closed orbit at $s_0$
  \State $\vec X_u$ \Comment{vector along the ray; $|\vec X_u|$ determines resolution}
  \State $a \gets 0, b \gets 2^S$ \Comment{low and high bounds on the ray}
\While{$b-a>1$}
  \State $m \gets (a + b) / 2$ 
  \State $\vec X \gets \vec X_0 + m \cdot \vec X_u$
  \State particle\_survived $\gets$ \textbf{tracking subroutine}($\vec X$)
  \State \algorithmicif \; particle\_survived \algorithmicthen \; $a \gets m$ \algorithmicelse \; $b \gets m$
\EndWhile
\State $\vec X_\text{aperture} \gets \vec X_0 + a \cdot \vec X_u$ \Comment{conservative estimate}
\EndFor
\end{algorithmic}
\caption{Binary search for dynamic aperture with $S$ search steps.\label{alg:da_binary}}
\end{algorithm}

Another approach to compute DA is grid probing (GP), where the $x$--$y$ plane is discretized into pixels. To perform an analysis of DA, one may perform tracking taking the center of each pixel as the initial coordinate, and then label it as ``captured'' or ``lost,'' thereby obtaining a rasterized image of the DA. Although it is computationally expensive, GP is capable of analyzing a DA with complex shapes, which may easily be missed by binary search.

\subsection{Flood fill}
\label{sec:floodfill}
\begin{figure}
    \centering
    \subfloat[]{ \includegraphics[width=0.7\columnwidth]{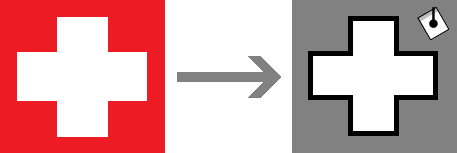}}
    \vspace{5pt}
    \subfloat[]{ \includegraphics[width=0.7\columnwidth]{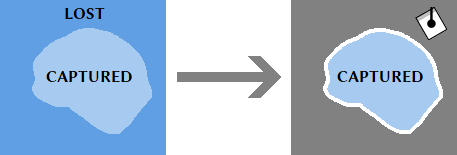}}
    \caption{(a) Flood-fill tool in raster graphics and (b) for DA computation. The algorithm will only query pixels in the respective connected area, and the boundary (black rim of cross and white corner of captured area, respectively).}
    \label{fig:ff-principle}
\end{figure}

Grid probing is computationally expensive due to many ``captured'' pixels in the DA interior: these pixels require the maximum number of turns to be evaluated, while ``lost'' pixels typically often terminate within very few turns. This asymmetry was the inspiration for applying the flood-fill tool \cite{floodfill}, commonly used in raster graphics programs, in the context of DA computation. When using the tool, the color under the cursor (e.g., red) is stored. Then, the algorithm traverses the pixels of the grid, searching for pixels that fulfill the criterion ``color is red'' starting from the cursor position, and breaking the search when encountering pixels which do not fulfill the criterion (see Fig.~\ref{fig:ff-principle}).

For DA computation, the aforementioned criterion is replaced with the outcome of a tracking computation, ``particle is lost,'' that is performed when the algorithm queries the pixel state. These pixels are then colored with a target color (e.g., gray). Note that the algorithm will only query the region with lost particles and the border of the captured region before breaking the search, but not the captured region, which would be computationally expensive. Further details can be found in Appendix~\ref{app:floodfill}. 

\subsection{Reverse scan\label{sec:reversescan}}
The binary search can be replaced by a reverse scan (RS) that is based on a concept similar to the flood fill (FF) algorithm applied to the DA computation: In Alg.~\ref{alg:da_binary}, the scan of the while-loop is replaced by a simple scan starting at $m=2^S$ (not from $m=0$) and varied decrementally until the boundary is found. Only one long tracking is needed for each ray.

\subsection{Performance comparison in OPA}
The existing and proposed algorithms are benchmarked to compare their performance. The flood-fill algorithm is compared with grid probing (rectangular grid), while the reverse scan algorithm is compared with binary search (polar grid). A distinction between methods using rectangular and polar grids must be made to ensure a fair comparison by having identical density of the search space. All algorithms were implemented in OPA~4.051 \cite{opa4051} to make this comparison (see Sec.~\ref{sec:results}). Only grid probing, which is thorough but slow, was implemented in the older versions. 

\begin{figure*}
\centering
\subfloat[Grid probing, $\dpp = 0\%$]{\includegraphics[width=0.45\textwidth]{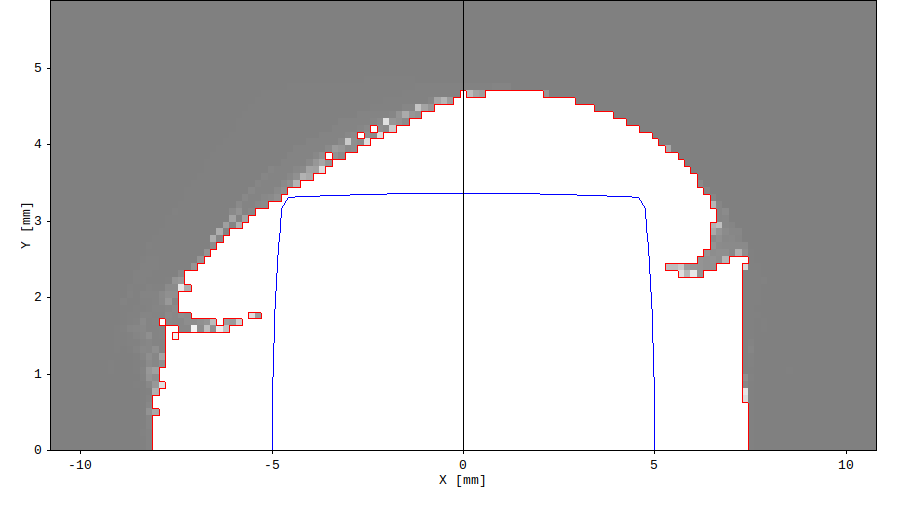}\label{fig:pda_a}}
\subfloat[Grid probing, $\dpp = 6\%$]{\includegraphics[width=0.45\textwidth]{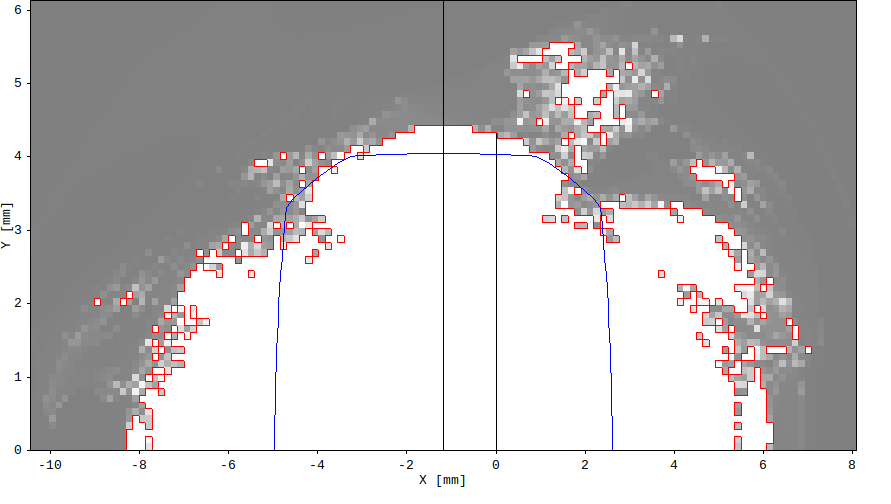}\label{fig:pda_b}}
\vspace{1pt}
\subfloat[Flood-fill, $\dpp = 0\%$]{\includegraphics[width=0.45\textwidth]{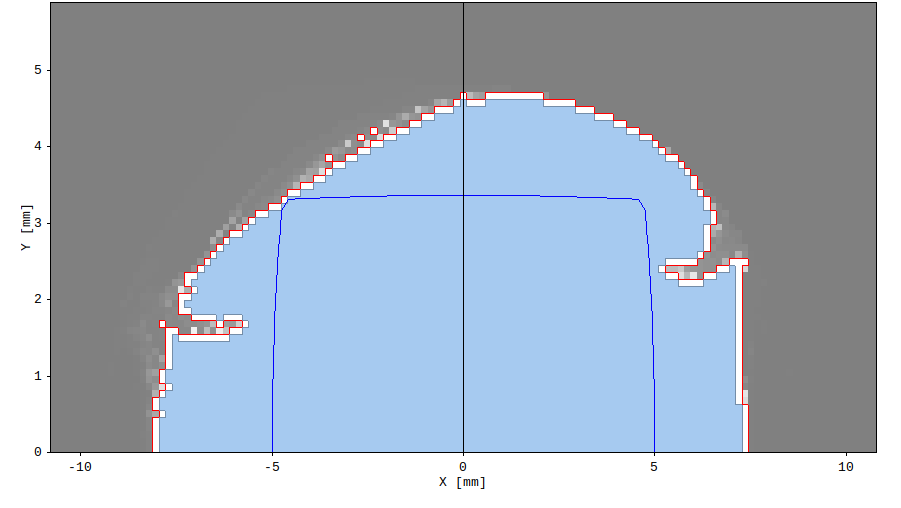}\label{fig:pda_c}}
\subfloat[Flood-fill, $\dpp = 6\%$]{\includegraphics[width=0.45\textwidth]{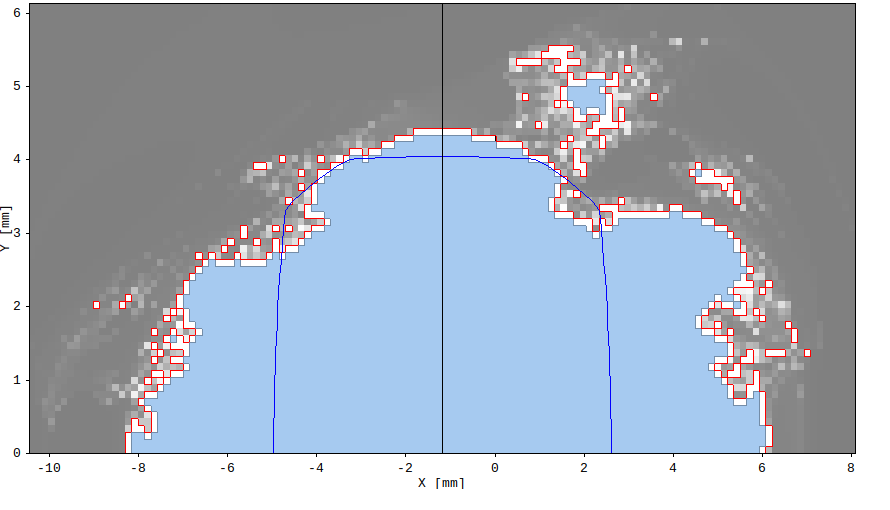}\label{fig:pda_d}}
\vspace{1pt}
\subfloat[Binary search, $\dpp = 0\%$]{\includegraphics[width=0.45\textwidth]{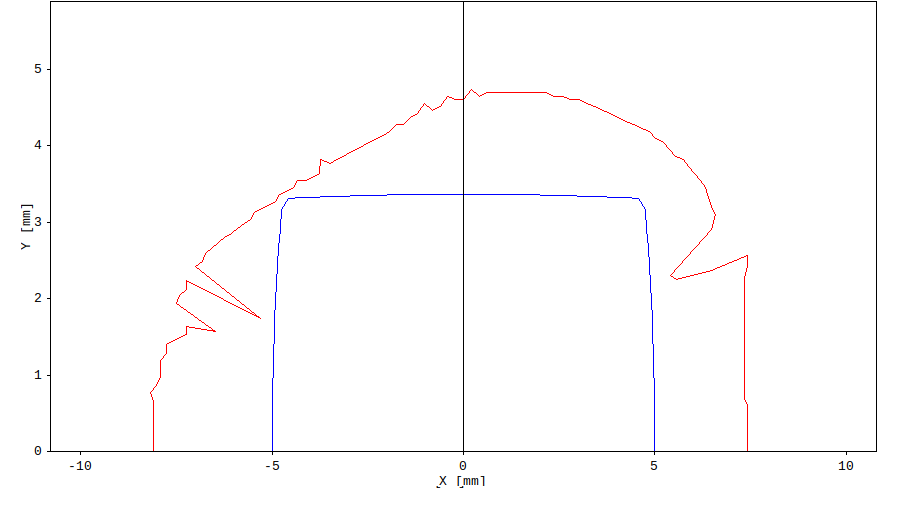}\label{fig:pda_e}}
\subfloat[Binary search, $\dpp = 6\%$]{\includegraphics[width=0.45\textwidth]{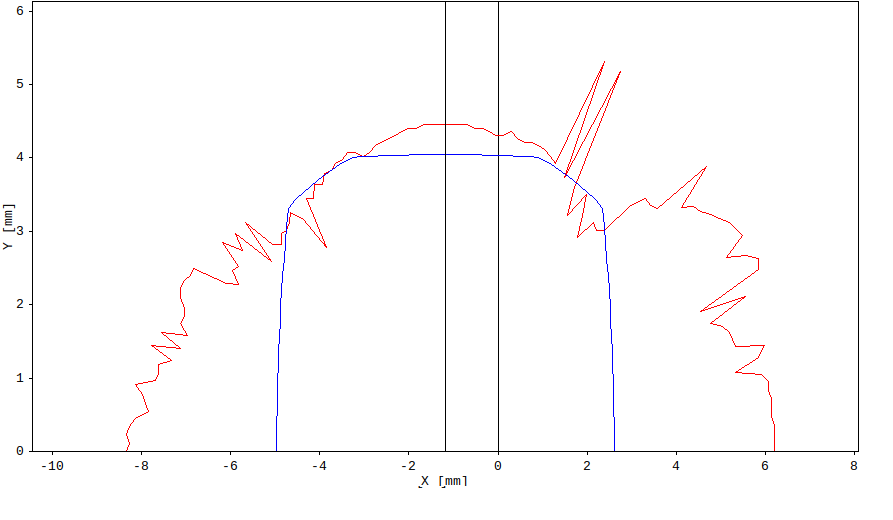}\label{fig:pda_f}}
\vspace{1pt}
\subfloat[Reverse scan, $\dpp = 0\%$]{\includegraphics[width=0.45\textwidth]{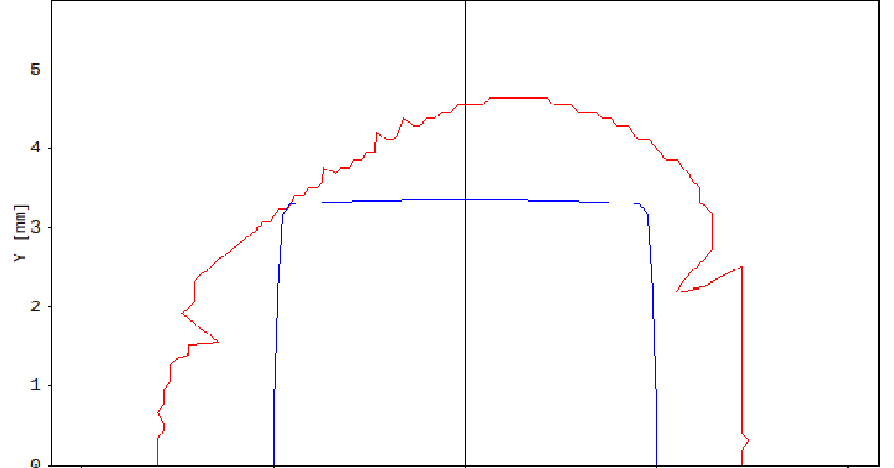}\label{fig:pda_g}}
\subfloat[Reverse scan, $\dpp = 6\%$]{\includegraphics[width=0.45\textwidth]{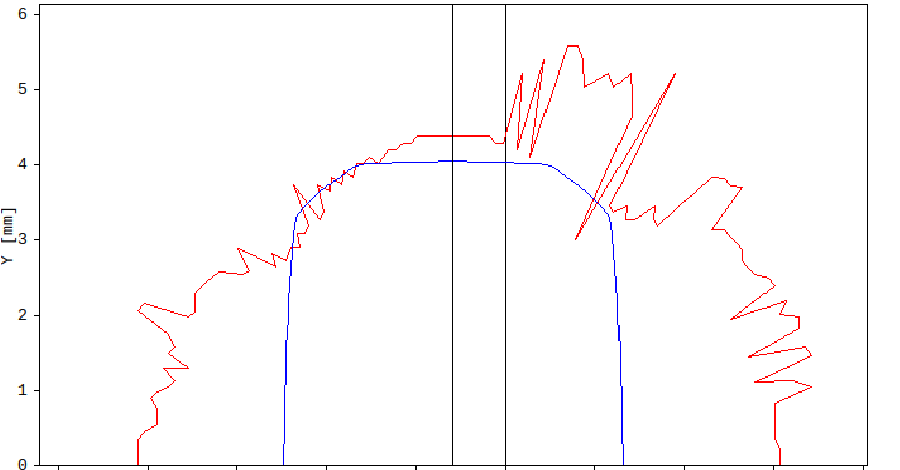}\label{fig:pda_h}}
\caption{Dynamic aperture of the SLS 2.0 lattice as computed in OPA for grid probing,   flood-fill, binary search and reverse scan methods in the upper $x,y$ half-plane for $N=500$ turns. The pixel images have a resolution of $129 \times 65$. The number of search rays is 129 with a resolution $<\num{8e-3}$. Left: on-momentum. Right: $\dpp=6\%$. Blue lines in these plots represent the physical apertures projected onto the $x$--$y$ plane of the reference location, where these DA are computed. They are ignored in the tracking.}
\label{fig:pda} 
\end{figure*}

Figure~\ref{fig:pda} shows OPA screenshots for the four methods of DA calculations for two cases of a smooth and a complex DA shape, which are obtained from the SLS 2.0 lattice~\cite{PhysRevAccelBeams.26.091601} on-momentum and at $+6\%$ momentum deviation. In Figs.~\ref{fig:pda_c} and~\ref{fig:pda_d} (flood fill), cells not tested are marked in blue.

A grid of $129\times 65$ points is used for GP or FF while, a relative resolution of 0.015 and a number rays of 129 for BS and RS. If a particle is not lost within 500 turns, it is considered stable.

Table~\ref{tab:megaturns} shows performance indicators for the compared methods; our figure of merit is the number of tracked turns required. For the smooth DA, the results indicate that FF is about 16 times faster than GP, while for the complex DA only a factor 6 is gained. RS is better than BS by a factor 3.3 and 2.5 for the smooth and complex DA, respectively. 

It is a prerequisite for the binary search that the array to be examined is sorted beforehand. This prerequisite is often not fulfilled. In Fig.~\ref{fig:pda}, the off-momentum dynamic aperture (right column) has isolated stable areas. The application of the binary search is therefore not fully justified. Identification of such islands fails also with the reverse search. The flood-fill algorithm can avoid this issue as demonstrated in Fig.~\ref{fig:pda_d}.

\begin{table}[!ht]
	\caption{\label{tab:megaturns}
		Total numbers of ``mega-turns'' ($\powten{6}$ turns) to compute dynamic apertures for on- and off-momentum (+6\%) particles using four different algorithms. In cases where all grid points are searched and stable, the total number of turns will be approximately 4.2 million in both rectangular (GP and FF) and polar grids (BS and RS).}
    \centering
    \begin{ruledtabular}
    \begin{tabular}{lcrr}
    & $\dpp$ & 0\% & 6\% \\
    \toprule
    Grid probing & {\bf GP} & 1.920 & 1.898 \\
    Flood fill & {\bf FF} & 0.122 & 0.316 \\
    \midrule    
    Binary search & {\bf BS} & 0.295 & 0.332 \\
    Reverse scan & {\bf RS} & 0.089 & 0.134 \\
    \end{tabular}
\end{ruledtabular}
\end{table}

\section{Touschek lifetime}
\label{sec:TL}
The calculation of the Touschek lifetime can be written as an integral involving optical functions and maximum and minimum stable relative momentum acceptances, $\min \dpp(s)$ and $\max \dpp(s)$, which both depend on the longitudinal position $s$~\cite{Bruck:1974ffz,piwinski1999touschek}.

Sampling of the integral therefore has to resolve the changes of bunch density, and this requires centimeter scale resolution. The calculation of the density from optical functions is, however, straightforward and computationally inexpensive. Hence the major part of the task at hand is the estimation of the local MA along the ring with sufficient sampling. It is best performed through particle tracking

\subsection{Standard approach}\label{sec:MAstandard}
The straightforward approach to compute local MA by tracking is the following \cite{sls1897a}:
\begin{quote}
  Will a pair of particles starting at location $s$ with coordinates $(x, x', y, y', \dpp) = (0, 0, 0, 0,\pm\dpp)$ survive (for a sufficient number of turns $N$) or not?
\end{quote}
Here an ideal (i.e., decoupled and error-free) lattice is assumed where the (on-momentum) closed orbit and the coordinate system coincide, and the angles $x'=p_x/p_0$ were used as coordinates. The number of turns $N$ should be on the order of the synchrotron oscillation period or the damping time, which is usually about 1000 turns or more. Thus Touschek lifetime calculations are computationally expensive in large part due to the required retrieval of local momentum acceptances. In example runs for the SLS 2.0 lattice, computation of the momentum acceptances
takes up about 90\% of the total computation time.

The minimum and maximum MAs are usually estimated via binary search, as outlined in Alg.~\ref{alg:touschek_binary}, but occasionally also using line search, as summarized in Appendix~\ref{app:linesearchMA}.

\begin{algorithm}
\begin{algorithmic}
\For{every relevant position $s_p$}
  \State $\vec X_0 \gets$ particle coordinates of closed orbit at $s_p$
  \State $a \gets 0, b \gets 2^S$ \Comment{low and high bounds}
\While{$b-a>1$}
  \State $m \gets (a + b) / 2$ 
  \State $\vec X \gets \vec X_0 +$ momentum offset $m \cdot \delta$ 
  \State particle\_survived $\gets$ \textbf{tracking subroutine}($\vec X$)
  \State \algorithmicif \; particle\_survived \algorithmicthen \; $a \gets m$ \algorithmicelse \; $b \gets m$
\EndWhile
\State $\dpp_\text{min/max}(s_p) \gets a\delta$ \Comment{conservative estimate}
\EndFor
\end{algorithmic}
\caption{\label{alg:touschek_binary} 
Local momentum acceptance computation for Touschek lifetime with $S$ search steps. $\dpp$ may be positive or negative depending on the sign of step size $\delta$.}
\end{algorithm}

\subsection{Fast Touschek Tracking (FTT)}\label{sec:interface}
The computationally expensive part in Alg.~\ref{alg:touschek_binary} is the tracking subroutine, which is also repeated many times. The process can, however, be significantly shortened by applying a more efficient estimator, while keeping the rest of the algorithm unchanged.
We imagine a particle that originated on the closed orbit, but obtained a momentum offset at a given position $s$. The particle is not to be tracked until it is lost, but only up to the first passage of a reference position $s_\mathrm{ref}$ in the lattice, where it enters with (6d) coordinates $\vec X$.

From this reference position, we could resume the tracking using $\vec X$ as starting coordinates to discover if the particle will survive; this would be standard Touschek tracking. But asking this is exactly equivalent to asking if $\vec X$ is located inside the (6d) dynamic aperture at location $s_\mathrm{ref}$ -- indeed, a direct check results in the same computation for both cases. 
It takes at most one turn of tracking for each particle to arrive at the reference position, $s_\mathrm{ref}$ -- the computational effort has been transferred into computation of the 6d aperture at $s_\mathrm{ref}$, which, without further constraints, is a computationally expensive task.  Fortunately, there are two reasonable approximations to reduce the number of dimensions in which the aperture needs to be computed:

\begin{enumerate}
\item We assume that the rf phase (or $ct$) shift of the particle during one turn is negligible relative to the extent of synchrotron motion. This approximation is good when the synchrotron frequencies are sufficiently small. With a constant $ct$ value for all particles at the reference plane, only a 5-dimensional aperture remains to be computed.
\item If the lattice is fully decoupled, then $\vec X$ will have $y,y'=0$. In the following we will focus on this case, as this reduces the aperture computation to the 3-dimensional space $x,x',\dpp$ the effect of coupling will be discussed in Sec.~\ref{sec:CoupledLattice}).
\end{enumerate}
With these approximations, we can now split the computation in two parts:

\paragraph{FTT precomputation:} 
Find a reasonable approximation of the DA at $s_\text{ref}$ in the three-dimensional space $x,x',\dpp$. In the following, this three-dimensional volume is referred to as the polyhedron.

\begin{figure*}[!ht]
    \subfloat[Standard tracking]{\includegraphics[width=0.45\textwidth]{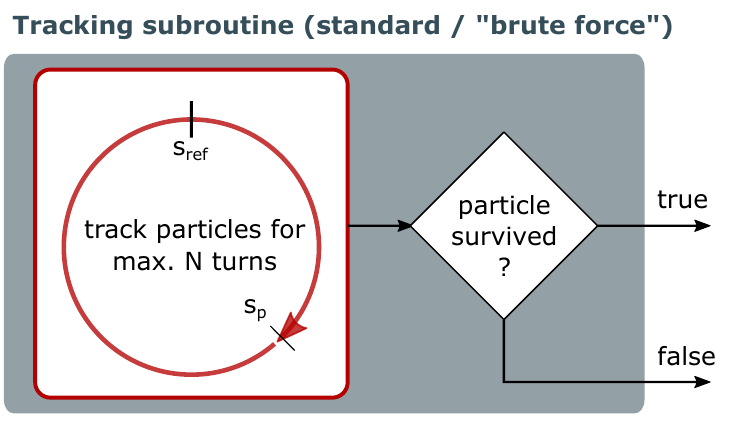}}
    \hspace{5mm}
    \subfloat[Fast Touschek Tracking]{\includegraphics[width=0.45\textwidth]{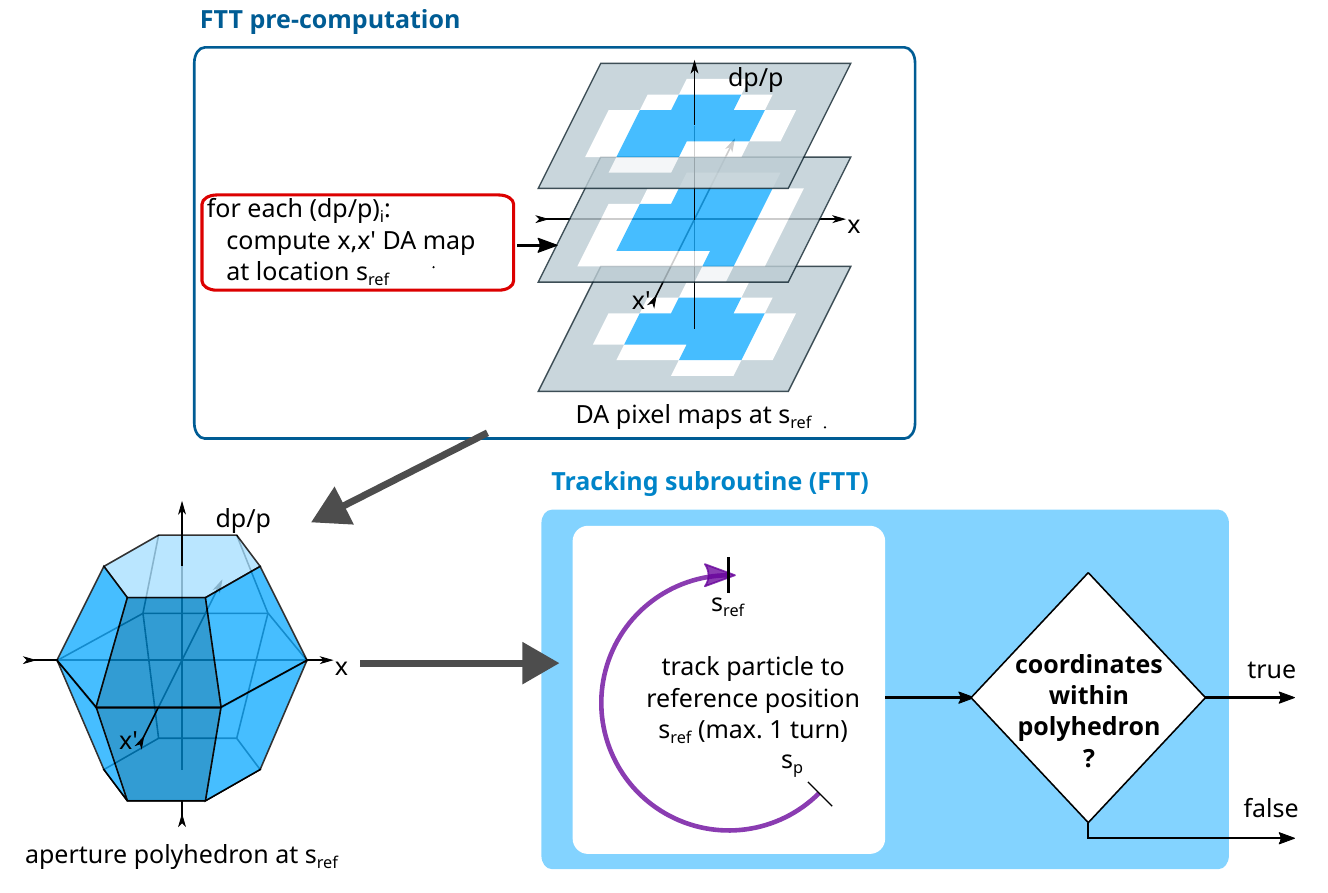}\label{fig:ftt-algo-overview_FTT}}
    \caption{\label{fig:ftt-algo-overview}Schematic view of a) standard tracking subroutine and b) Fast Touschek Tracking which utilizes the interface approach. The aperture polyhedron is typically asymmetric.}
\end{figure*}

\paragraph{FTT tracking subroutine:} 
Using the polyhedron at location $s_\mathrm{ref}$ as an interface, we only need to track particles up to the reference position -- that is, for at most one turn -- to see if they are captured or lost, i.e., if inside the polyhedron or not. The outline of the resulting FTT algorithm is visualized in Fig.~\ref{fig:ftt-algo-overview}. 

In the following, we go into the details of precomputation and introduce polygon grid stepping and the region checking (particle coordinates in polyhedron).

\subsubsection{FTT precomputation}
\label{sec:precompute}
To compute the aforementioned aperture polyhedron, our initial version of FTT utilizes equidistant DA slices, each with constant $\dpp$ value.
The step size between the slices can be set completely independently of the step size $\delta$ in the Touschek algorithm (Alg.~\ref{alg:touschek_binary}) and, in practice, significantly larger. The aperture in each slice is computed as a polygon, and the polygons of different $\dpp$ slices are later connected to form the polyhedron (see Fig.~\ref{fig:ftt-algo-overview_FTT}).

\begin{figure}
  \centering
  \begin{tikzpicture}
  \fill[black!30!white] (0,0) rectangle (3.2,3.2);
  \fill[white] (2.0,0.0) rectangle (2.4,0.4);
  \fill[white] (0.4,0.4) rectangle (2.8,2.0);
  \fill[white] (0.8,2.0) rectangle (2.8,2.4);   
  \fill[white] (1.2,2.0) rectangle (2.4,2.8);
  \fill[uncomputed] (2.0,0.4) rectangle (2.4,0.8);
  \fill[uncomputed] (0.8,0.8) rectangle (2.4,2.0);  
  \fill[uncomputed] (1.2,1.6) rectangle (2.4,2.4);
  
  \coordinate (orig) at (1.6,1.6);
  \coordinate (u) at (1.6,2.8);
  \coordinate (uuur) at (1.9,2.8);
  \coordinate (uur) at (2.2,2.8);
  \coordinate (ur) at (2.4,2.4);
  \coordinate (urr) at (2.8,2.2);
  \coordinate (urrr) at (2.8,1.9);
  \coordinate (r) at (2.8,1.6);
  \coordinate (drrr) at (2.8,1.3);
  \coordinate (drr) at (2.8,1.0); 
  \coordinate (dr) at (2.8,0.4);
  \coordinate (ddr) at (2.4,0.0);
  \coordinate (dddr) at (2.0,0.0);
  \coordinate (d) at (1.6,0.4);
  \coordinate (dddl) at (1.3,0.4);
  \coordinate (ddl) at (1.0,0.4);
  \coordinate (dl) at (0.4,0.4);
  \coordinate (dll) at (0.4,1.0);
  \coordinate (dlll) at (0.4,1.3);  
  \coordinate (l) at (0.4,1.6);
  \coordinate (ull) at (0.8,2.0); 
  \coordinate (ulll) at (0.4,1.9);
  \coordinate (ul) at (0.8,2.4); 
  \coordinate (uul) at (1.2,2.4);
  \coordinate (uuul) at (1.3,2.8);
  
  \draw[thick] (u) -- (uuur) -- (uur) -- (ur) -- (urr) -- (urrr) -- (r) -- (drrr) -- (drr) -- (dr) -- (ddr) -- (dddr) -- (d) -- (dddl) -- (ddl) -- (dl) -- (dll) -- (dlll) -- (l) -- (ulll) -- (ull) -- (ul) -- (uul) -- (uuul) -- cycle;
  \draw[gray] (u) -- (orig) -- (d);
  \draw[gray] (uuur) -- (orig) -- (dddl);  
  \draw[gray] (uur) -- (orig) -- (ddl);  
  \draw[gray] (ur) -- (orig) -- (dl);
  \draw[gray] (urr) -- (orig) -- (dll);
  \draw[gray] (urrr) -- (orig) -- (dlll);
  \draw[gray] (r) -- (orig) -- (l);
  \draw[gray] (drr) -- (orig) -- (ull);
  \draw[gray] (drrr) -- (orig) -- (ulll);
  \draw[gray] (dr) -- (orig) -- (ul);
  \draw[gray] (ddr) -- (orig) -- (uul);
  \draw[gray] (dddr) -- (orig) -- (uuul); 
  \end{tikzpicture}
  \caption{\label{fig:polygon_stepper}
    Conversion of DA in a rectangular pixel grid into polygon using grid stepping and the 2d fixpoint. Note that we start from the center seeking outwards. Also, the computation is inexpensive as the tracking has already been performed.}
\end{figure}
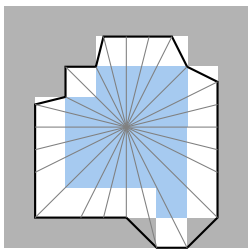

The $x$--$x'$ DA for a given $\dpp$-slice can be found using any DA algorithm, e.g., flood-fill from Sec.~\ref{sec:floodfill} or binary search. After calculating the DA for each $\dpp$-slice, it is beneficial to have a common discretization because the rectangular grid is not convenient, as will be evident later. To create this in each slice, rays are constructed in equidistant increasing angles $\theta_k$ originating from the fixed point (FP) , i.e., the closed orbit, for the given $\dpp$ plane $i$, and the radius $r_{i,k}$ of each ray is increased (starting from zero) so that the point
\begin{equation}
    \begin{pmatrix}
    x \\ x'
    \end{pmatrix}_{i,k}
    =
    \begin{pmatrix}
    x \\ x'
    \end{pmatrix}_{i,\mathrm{FP}}
    + r_{i,k}
    \begin{pmatrix}
    \cos \theta_k \\ \sin \theta_k    
    \end{pmatrix}
    \label{eq:inplane_polygon}
\end{equation}
lies on the boundary between a captured and lost pixel in the given plane $i$ (see Fig.~\ref{fig:polygon_stepper}).

The points with indices $i,k$ at 
\begin{equation*}
\begin{pmatrix} x_{i,k} \\ x'_{i,k} \\ \dpp_i \end{pmatrix}
\end{equation*}
then are the boundary points of the polyhedron. Compared to tracking, this procedure is numerically inexpensive, and provides a convenient mesh points $(i,k)$.

\subsubsection{FTT tracking subroutine}
Having obtained the DA polyhedron, we may now perform the modified tracking subroutine (of Alg.~\ref{alg:touschek_binary}), which consists of two steps:
\begin{enumerate}
\item track the particle from the location of its Touschek event to the reference position $s_\text{ref}$ (which is always less than one turn around the ring) 
\item check if the particle with coordinates $(x,x',\dpp)$ is located inside the DA polyhedron or not.
\end{enumerate}

The checking procedure is a variant of the point-in-polygon problem. In our case, it can be reduced from 3d to 2d by linear interpolation 
between the two polygons $i$ and $i+1$ closest to the $\dpp$ value of the point to be checked. Geometrically, this is equivalent to ``slicing'' the polyhedron with a plane of constant $\dpp$. The points of the interpolated polygon are then given by 
\begin{equation}
\begin{split}
    \begin{pmatrix}
    x \\ x'
    \end{pmatrix}_k
    &=
    (1-\lambda) 
    \begin{pmatrix}
    x \\ x'
    \end{pmatrix}_{i,k} 
    + \lambda \begin{pmatrix}
    x \\ x'
    \end{pmatrix}_{i+1,k} \\
    \mathrm{with} \quad \lambda &= \frac{\dpp-\dpp_i}{\dpp_{i+1}-\dpp_i}. 
    \end{split}
    \label{eq:xxp_interpol}
\end{equation}
The same interpolation rule holds for the interpolated fixpoint. The above interpolation would be much more complicated with the DA slices of the rectangular grid. Using the interpolation, we can obtain the angle
\begin{align}
    \phi &= \mathrm{atan2}(x'- x'_\mathrm{FP}, x - x_\mathrm{FP}), \label{eq:phi_interpol}
\end{align}
which allows finding the two adjacent rays and their angles $\theta_k$ and $\theta_{k+1}$ values even before computing Eq.~\ref{eq:xxp_interpol}. The condition for acceptance, meaning that the particle is captured, follows from elementary geometry~\footnote{a cross-product is used to check if the point in question is below the connection line of the 2d vectors $\vec x_k$ and $\vec x_{k+1}$}:
\begin{equation}
\begin{split}
x'\Delta x-x\Delta x' &> x'_k \Delta x - x_k \Delta x' \\ \mathrm{with} \quad
\Delta x = x_{k+1}-x_k&, \quad \Delta x' = x'_{k+1}-x'_k. 
\end{split} \label{eq:linesegment}
\end{equation}

\subsection{Summary of Fast Touschek Tracking algorithm}
\label{sec:ftt}
Having all necessary subroutines of the Fast Touschek tracking algorithm at hand, we now summarize it as follows (compare Fig.~\ref{fig:ftt-algo-overview}):
\begin{enumerate}
\item Perform a calculation of dynamic aperture in the $x$--$x'$ plane at one particular location $s_\text{ref}$ (e.g., $s_\text{ref}=0$) for a set of momenta $\dpp_i$. This defines a 3-dimensional acceptance volume $(x,x',\dpp)$, i.e., the interface polyhedron.
\item Track a particle starting at the on-momentum closed orbit but with $\dpp$ momentum offset, from an $s$-location to $s_\text{ref}$. 
\item For this value of $\dpp$ take a cross-section through the 3-dimensional acceptance volume by interpolation to create an interpolated polygon.
\item Evaluate whether $x$--$x'$ coordinates of tracked particle are within the polygon.
\item Repeat steps 2--4 for various $\dpp$ values to search the local MA with sufficient resolution. 
\item Repeat step 5 for each $s$-location to obtain the MA along the ring.
\end{enumerate}
Computing the dynamic apertures leading to the polyhedron is the most time consuming step of the FTT procedure. Applying fast algorithms like flood-fill or reverse scan is beneficial. The second step is fast, since tracking is done for maximum one full turn per particle (this tracking can in principle be limited to maximum half a turn by using back-tracking).

For the fifth step, instead of having a discrete array of $\dpp$-values, search algorithms (e.g., binary-search or reverse-scan) can be applied to find the maximum/minimum of the local momentum acceptance. The algorithm for such a binary search implementation is shown in Alg.~\ref{alg:ma_ftt_binary}.

\begin{algorithm}
\begin{algorithmic}
\State $\mathbf{V} \gets$ precomputed polyhedron in $x$--$x'$--$\dpp$ space
\For{every relevant position $s_p$}
  \State $\vec X_0 \gets$ closed orbit at $s_p$
  \State $a \gets 0, b \gets 2^S$ \Comment{low and high bounds}
\While{$b-a>1$}
  \State $m \gets (a + b) / 2$ 
  \State $\vec X \gets \vec X_0 +$ momentum offset $m \cdot \delta$ 
  \State $\vec X_\text{ref}$ $\gets$ tracking $\vec X$ to location $s_\text{ref}$
  \State particle\_in\_polyhedron $\gets$ $\vec X_\text{ref} \in \mathbf{V}$
  \State \algorithmicif \; particle\_in\_polyhedron \algorithmicthen \; $a \gets m$ \algorithmicelse \; $b \gets m$
\EndWhile
\State $\dpp_\text{min/max}(s_p) \gets a\delta$ \Comment{conservative estimate}
\EndFor
\end{algorithmic}
\caption{\label{alg:ma_ftt_binary} 
Local momentum acceptance computation found using FTT with binary search with $S$ search steps. $\dpp$ may be positive or negative depending on the sign of step size $\delta$.}
\end{algorithm}

We assume the tracking effort for one turn in a third or next-generation light-source storage ring to scale roughly proportional to the number of dipoles, $N_d$. In third and next-generation light-source storage rings, the beam is focused at each dipole to achieve low emittance. The number of elements or the number of tracking steps is therefore approximately proportional to $N_d$. The standard Touschek tracking requires this tracking to be performed for obtaining the MA for each location in the ring. The number of $s$-locations to resolve the variation of optical functions is proportional to $N_d$ since, again, the beam is focused at each dipole. With these observations we see that the computational effort to obtain Touschek lifetime roughly scales with $N_d^2$. $N_d$ can be replaced by $Q_x$ when we compare the lattices where the horizontal phase advances between dipoles are comparable. In the above scaling we did not include the number of turns in the local MA computation; the synchrotron tune and the damping time may vary from ring to ring. The factor $N_d^2$ may be, however, dominant compared to this variation.

Due to the interface approach utilized by FTT, the computational effort is concentrated in finding the 3d dynamic aperture at the reference location. Thus, the effort to find the MA with FTT roughly scales with $N_d$. FTT is therefore interesting especially for the ultralow-emittance ring with a very large number of dipoles, including those for which the standard momentum acceptance tracking comes with such high effort that it is impractical.

\section{FTT implementation into accelerator codes}
\label{sec:results}
The FTT procedure has been implemented into two common tracking codes: OPA~\cite{opa4051} and Accelerator Toolbox (AT)~\cite{terebilo_2001,atcollab}. The methods are now publicly available: in OPA as part of the version 4.051 release, while the AT-MATLAB implementation with examples can be accessed on GitHub~\cite{FTTgithub}. The tracking in OPA is four-dimensional with $\dpp$ being finite but constant, while AT offers six-dimensional tracking, including synchrotron radiation damping. In the following, we will present results from the two implementations, while increasing the complexity of the lattices to be evaluated using FTT.

In the tracking results presented in this section, the physical apertures were included. We still refer to the stable area in the $x$--$x'$ plane as ``dynamic aperture (DA)'' for simplicity and readability but the acceptances are limited by both dynamics and physical objects. 

\subsection{OPA}
\begin{figure*}
  \subfloat[$x$-$x'$-$\dpp$ DA polyhedron projected on the $x$--$x'$ plane.
    Variation of momentum offset is represented by the color: negative to positive $\dpp$
    as blue to red. ]{\includegraphics[width=0.45\textwidth,valign=c]{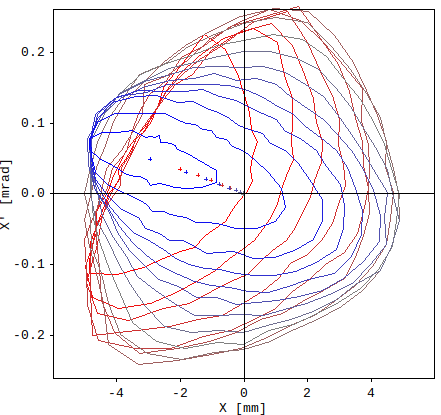}\label{fig:OPApolyhedron}}
    \hspace{2mm}
    \subfloat[Example of particles from various values of $s$ tracked to $s_\text{ref}$ for a given value of $\dpp$.]{\includegraphics[width=0.45\textwidth,valign=c]{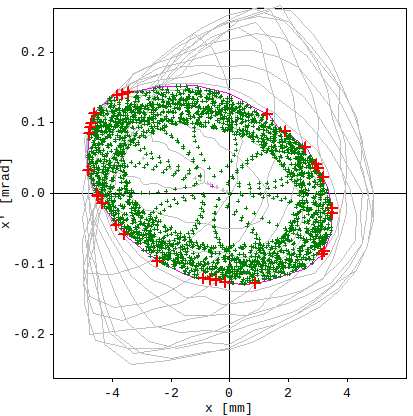}\label{fig:OPAinterpolated}}
    \vspace{1pt}
    \subfloat[Example of momentum acceptance computed with standard tracking (red) and FFT (magenta). The rf bucket is indicated by green lines while the MA in brown corresponds to the physical aperture.]{\includegraphics[width=0.45\textwidth,valign=c]{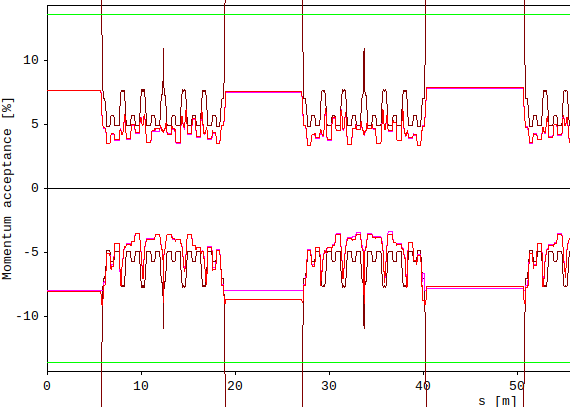}\label{fig:OPAmomentumAcceptance}}
    \caption{\label{fig:ftt}Momentum acceptance computed using Fast Touschek Tracking for the SLS 2.0 lattice without coupling, insertion devices and machine imperfections. Flood-fill is used with a 65$\times$65 grid cells, 17 $\dpp$ slices and 36 polygon points. 100 turns are tracked.}
\end{figure*}

The screenshots of Fig.~\ref{fig:ftt} demonstrate FTT for the SLS 2.0 lattice, as preliminarily implemented in OPA. The flood-fill algorithm is used to find the dynamic apertures for each $\dpp$-slice at the reference point.  Figure~\ref{fig:OPApolyhedron} shows the $x$--$x'$--$\dpp$ DA projected onto the $x$--$x'$ plane. The dots indicate the closed orbits for the various values of $\dpp$. 

Next, Fig.~\ref{fig:OPAinterpolated} shows (in magenta) an interpolated polygon for one $\dpp_i$ value. The red/green crosses indicate particles tracked for momentum offset $\dpp_i$ from various longitudinal positions $s_k$ to the references position $s_\text{ref}$. Green crosses indicate the particles that are within the polyhedron, while red crosses are outside, and will be lost if further tracked. 

Finally, Fig.~\ref{fig:OPAmomentumAcceptance} compares the results for the local momentum acceptances obtained from standard tracking red lines) and from the FTT procedure (magenta). Touschek lifetime results are in a good agreement: 3.31~h with FFT vs. 3.45~h with standard tracking. The brown curve in Fig.~\ref{fig:OPAmomentumAcceptance} is the linear acceptance derived from (on-momentum) beta functions and apertures, and the green line is the rf acceptance, which had been set to a large value here in order not to be the limiting acceptance.

\subsection{Accelerator Toolbox}\label{sec:FTT-AT}
\begin{figure}
    \centering
    \subfloat[]{\includegraphics[width=\columnwidth]{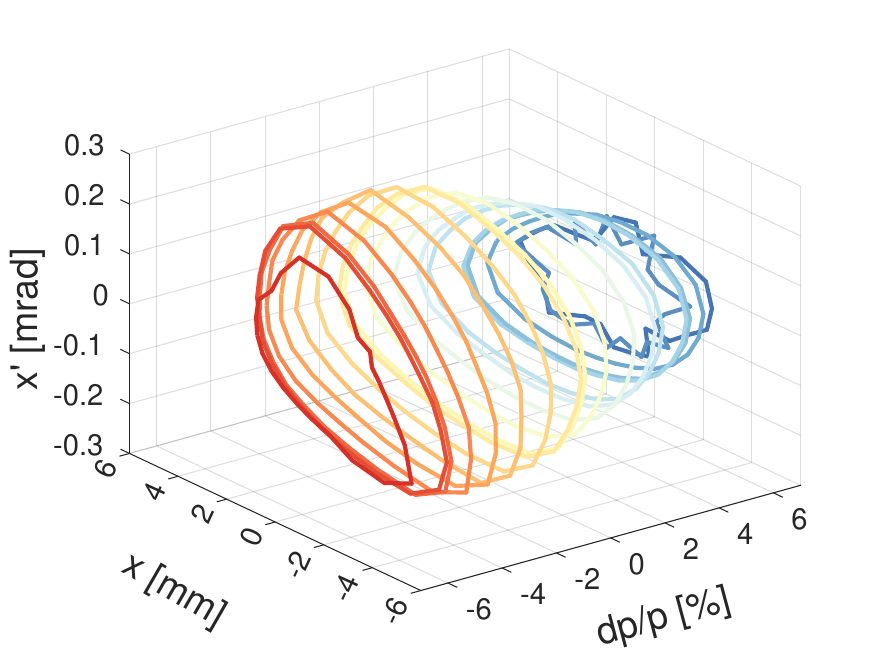}}
    \vspace{1pt}
    \subfloat[]{\includegraphics[width=\columnwidth]{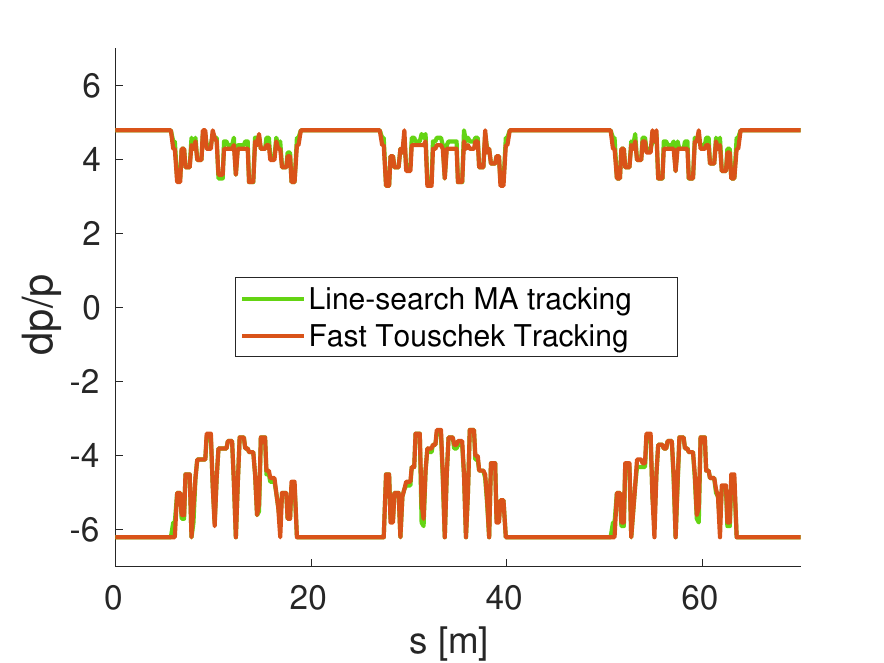}}
    \caption{\label{fig:ATMomentumAcceptanceExample} a) Computed $x$--$x'$--$\dpp$ dynamic aperture,
      b) Comparison between momentum acceptances computed in Accelerator Toolbox
      using the standard approach and the FTT method for the first three arcs of the ideal SLS 2.0 lattice.}
\end{figure}

To further benchmark the FTT method an implementation is made in the widely used tracking code Accelerator Toolbox~\cite{terebilo_2001,atcollab}. The SLS 2.0 lattice is again used for benchmarking: First, the MA is computed in the standard way using the line-search algorithm summarized in Appendix \ref{app:linesearchMA}) at the entrance of all lattice elements with non-zero length. Two other algorithms for finding MA are also used: binary search (Sec.~\ref{sec:MAstandard}) and sieve search (Appendix~\ref{app:sieve-search}). The latter algorithm is an optimized version of line search providing a factor two speed-up but giving exactly the same results. Second, the FTT algorithm is used, and the MAs obtained are compared with the standard tracking (line search) MA as shown in Fig.~\ref{fig:ATMomentumAcceptanceExample}. The $x$--$x'$ DA slices for various values of $\dpp$ computed using FTT are shown in Fig.~\ref{fig:ATMomentumAcceptanceExample}. For FTT, the algorithm used in the MA search is binary search. The DA slices are also calculated using the binary search algorithm (Alg.~\ref{alg:da_binary}). All tracking simulations include classical radiation damping (quantum excitation not included).

The calculated Touschek lifetime and total number of turns involved in obtaining these results are summarized in Table~\ref{tab:megaturns_MA}. The Accelerator Toolbox implementation of Piwinski's formula for the Touschek lifetime is used~\cite{piwinski1999touschek,nash_2015}. The FTT method requires almost 400 times fewer turns than the standard MA tracking with line search while providing very similar results. With respect to binary search, FTT uses around 36 times fewer number of turns.

The comparison indicates that the local MA does not have a clear, single boundary between ``capture'' and ``lost'' (similarly to the islands observed in DA) at some locations since the binary search lifetime is not fully consistent with the line/sieve search lifetime. The lifetime with line/sieve search is slightly underestimated, while it is slightly overestimated with binary search.

The MA within FTT is based on a binary search for each $s$-position. A line search was also tested, but it turned out that the MA, and hence the lifetime, from FTT did not depend on the internal MA search algorithm in this case. The shorter lifetime from FTT is therefore mainly related to the DA polyhedron. The lifetime from the FTT may tend to be underestimated because the surfaces of the polyhedron are mostly convex, and the polyhedron is confined by these surfaces. The issues in computing DA slices from the accuracy point of view are discussed in Appendix~\ref{AppendixIslands}. Nonetheless, the accuracy of the FTT lifetime for the ideal lattice is about 93\% (underestimation similar to OPA result), which is satisfactory for our lattice characterization purpose.

\begin{figure}
    \centering
    \subfloat[]{\includegraphics[width=0.5\columnwidth]{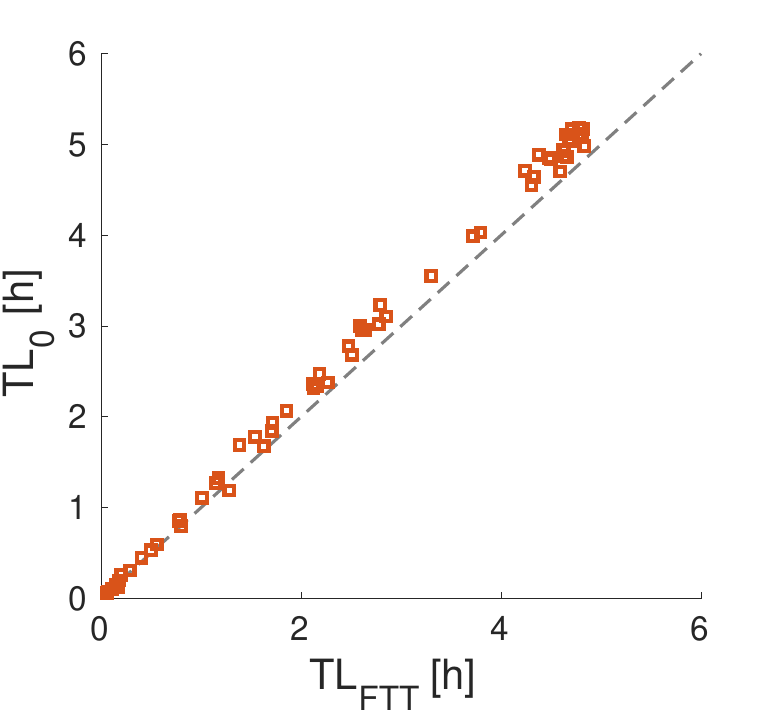}\label{fig:ATLifetimeComparison_plot}}
    \subfloat[]{\includegraphics[width=0.5\columnwidth]{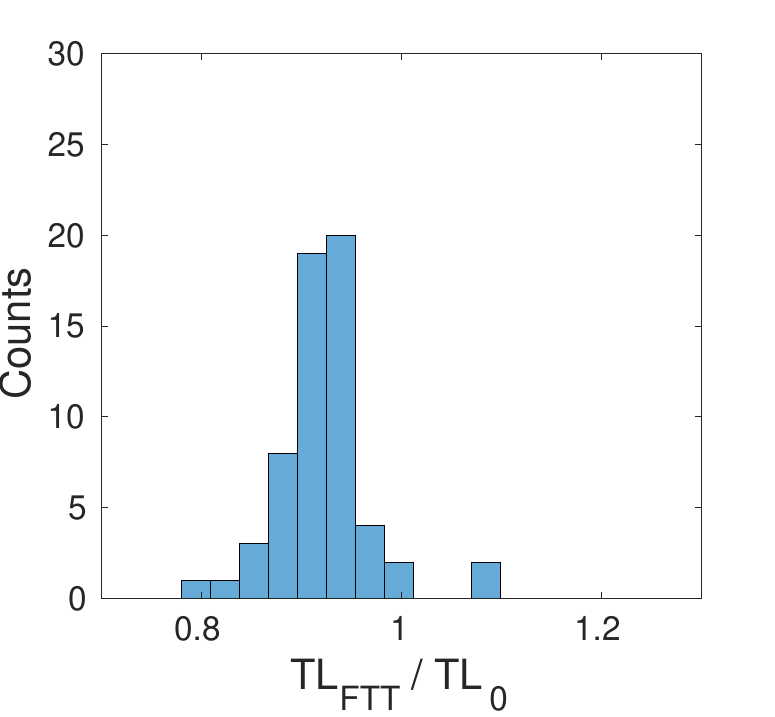}\label{fig:TLMATvsTLFTT_histogram}}
    \caption{\label{fig:ATLifetimeComparison}Comparison between Touschek lifetimes calculated from momentum acceptances computed using the standard tracking technique using the line-search algorithm (TL$_0$) and from Fast Touschek Tracking (TL$_\text{FTT}$). }
\end{figure}

To investigate the behavior of FTT over a broad range of lifetimes, sextupole errors of varying severity are introduced into the lattice. Sixty samples of sextupole errors are created, and the resulting lifetimes are evaluated. In Fig.~\ref{fig:ATLifetimeComparison_plot} the lifetimes from the standard tracking with line search are plotted against the ones from FFT for all sixty seeds, while Fig.~\ref{fig:TLMATvsTLFTT_histogram} shows a histogram of their ratios. In addition, the statistics of these results are summarized in Table~\ref{tab:TL_comparison}, including the results of standard tracking with binary search and sieve search. When the average lifetime over 60 seeds is considered, the accuracy of the FTT lifetime is similar to the case without sextupole errors (93\%). However, there are several cases where the FTT lifetime differs from the standard tracking by more than 7\%. We discuss this further in the next subsection. 

In Table~\ref{tab:TL_comparison}, we also list the statistics of the computation time (or CPU time) for each method, instead of the number of turns, as a more interesting figure of merit. The absolute computation time depends on the performance of the computer, and this is not our interest here, whereas the relative comparison can be regarded as a general conclusion: FTT can be significantly faster than standard tracking.

\begin{figure}
    \centering
    \includegraphics[width=1\columnwidth]{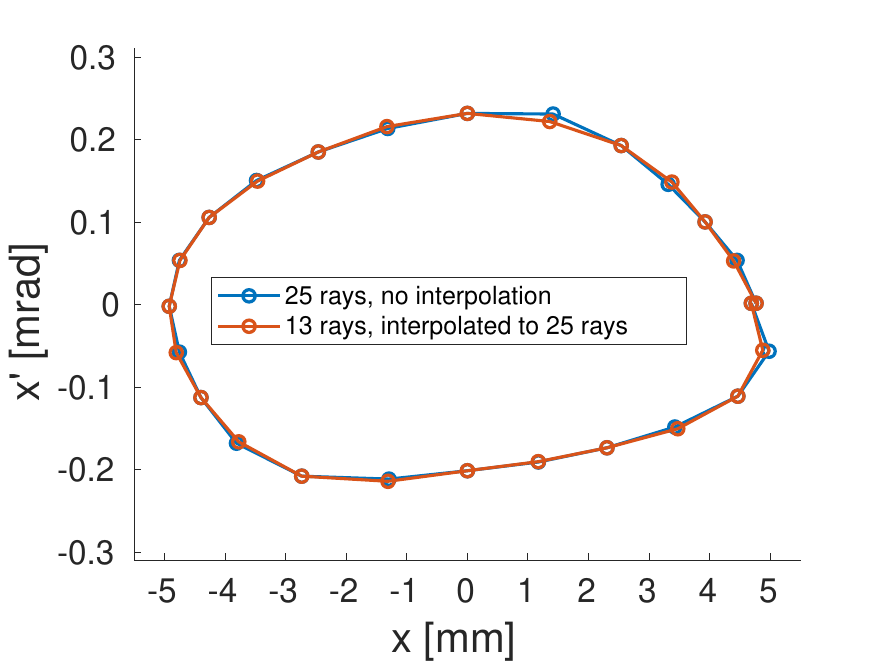}
    \caption{\label{fig:DAexample_interpolation}Dynamic apertures
      in $x$--$x'$ space for $\dpp=0$ computed using binary search with either 25 rays or 13 rays with interpolation. }
\end{figure}

It is possible to speed up the DA-slice computation by reducing the resolution, i.e., the number of rays in binary search. The number of rays can be effectively recovered by an interpolation between the rays, however. Figure~\ref{fig:DAexample_interpolation} is an example, where the number of rays in the binary search DA computation has been reduced from 25 to 13, but with a cubic interpolation being applied. Using this technique, the total CPU time is reduced by 45\% (since the majority of the CPU time comes from the DA calculation) with marginal loss in accuracy. The reverse scan algorithm (Sec.~\ref{sec:reversescan}) may be an alternative to binary search, and we would expect yet another speed-up of the FTT calculation. The choice of algorithms and parameters (step sizes or resolutions) needs to be optimized such that the required accuracy is fulfilled, and one may need some test runs for the lattice to be examined before launching a series of lifetime computations.

\begin{table}[!ht]
	\caption{Total numbers of turns for momentum acceptance tracking in the SLS 2.0 lattice
		using various search algorithms and calculated Touschek lifetime.
		The first three rows are results obtained from the standard MA tracking
		but with different algorithm for the MA search.
		The last one is from FTT with a binary MA search.
		No machine imperfections included.}
    \centering
    \begin{ruledtabular}
    \begin{tabular}{lcS[table-format=1.2]}
    & {Turns [$\times 10^6$]}& {TL [hours]}  \\
    \toprule
    Line search & 479 & 5.11 \\
    Binary search  & 41.1 & 5.18 \\
    Sieve search  & 216 & 5.11 \\
    Fast Touschek Tracking & 1.50 & 4.78  \\
    \end{tabular}
\end{ruledtabular}
    \label{tab:megaturns_MA}
\end{table}

\begin{table}[!ht]
	\caption{Average Touschek lifetime normalized
		to the lifetime from the standard tracking with line search (TL$_0$)
		and average CPU usage for various momentum acceptance algorithms.
		The computation time and the lifetime depend on the sextupole setting,
		their standard deviations over 60 seeds are also listed.}
    \centering
    \begin{ruledtabular}
    \begin{tabular}{lcc}
    & {CPU time [hours]}& {TL/TL$_0$}  \\
    \toprule
    Line search & $166.7 \pm 27.0$ & $1.00 \pm 0.00$ \\
    Binary search  & $23.7\pm 1.13$ & $1.03\pm 0.06$ \\
    Sieve search  & $84.2\pm 5.5$ & $1.00\pm 0.00$ \\
    Fast Touschek Tracking & $1.09\pm 0.15$ & $0.93\pm 0.07$  \\
    \end{tabular}
\end{ruledtabular}
    \label{tab:TL_comparison}
\end{table}

\subsection{Lattices with errors}
\begin{figure}
    \centering
    \subfloat[]{\includegraphics[width=0.5\columnwidth]{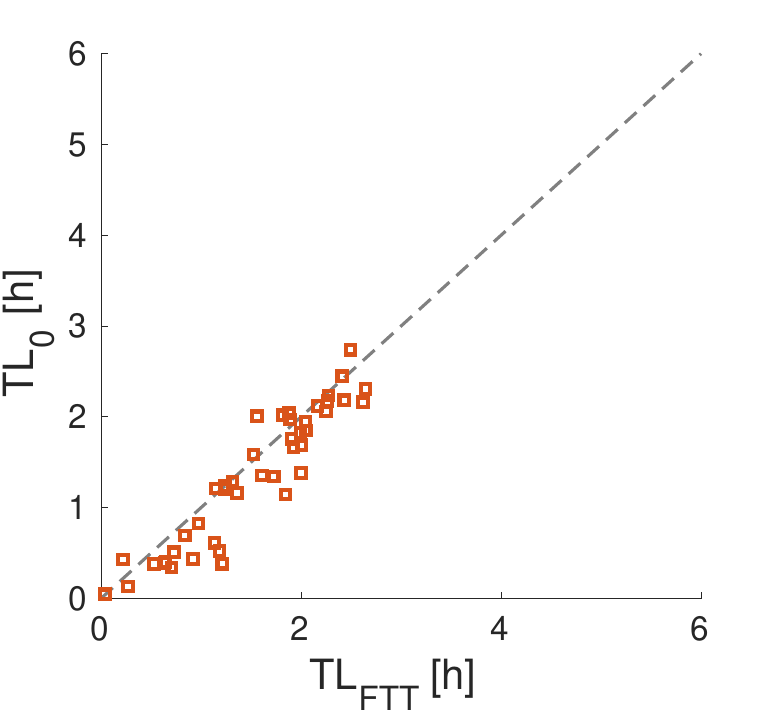}\label{fig:TLMATvsTLFTT_afterRFON}}
    \subfloat[]{\includegraphics[width=0.5\columnwidth]{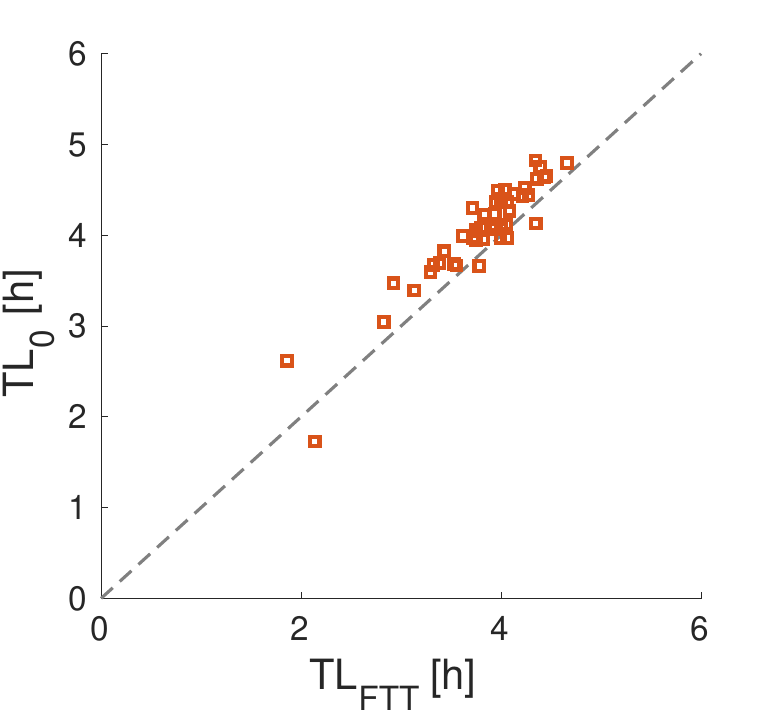}\label{fig:TLMATvsTLFTT_afterLOCO}}
    \caption{\label{fig:TLMATvsTLFTT_SC}Comparison between Touschek lifetimes calculated using the standard MA tracking technique and the Fast Touschek Tracking for SLS 2.0 lattices propagated through Simulated Commissioning to stages: a) after establishing closed orbit and initial orbit correction but without BBA or LOCO b) after BBA and LOCO. }
\end{figure}

The approximations that we made in Sec.~\ref{sec:interface} make the polyhedron in $x$--$x'$--$\dpp$ a valid interface to determine whether the Touschek scattered particle is within the MA: synchrotron motion is negligible over one turn, and the lattice is uncoupled. The latter aproximation, however, might not be valid in real machines with magnet misalignments and field errors, which introduce transverse coupling. The Simulated Commissioning toolkit of the Accelerator Toolbox has been used to generate 40 seeds with these errors and propagate them through the realistic commissioning process~\cite{hellert_sc}. Then, the momentum acceptance and Touschek lifetime are evaluated using both the standard MA tracking and FTT. The evaluation is done at two points of the commissioning process: a) after closed orbit has been established with orbit correction, but no beam-based alignment (BBA) nor optics correction applied, and b) after BBA and optics correction using Linear Optics from Closed Orbits (LOCO)~\cite{SAFRANEK199727}. The results are shown in Fig.~\ref{fig:TLMATvsTLFTT_SC}, separately for the above two cases.

The agreement between the two methods is not as good as the previous benchmarking, especially for the lattices before BBA and the optics correction. We looked into the DA slices of those cases and found that a ``fuzzy'' dynamic aperture as is also seen in the upper right plot of Fig.~\ref{fig:pda}), which appeared more often in the uncorrected lattice, could be a source of discrepancy rather than the transverse coupling. Nevertheless, the lifetime after the corrections is more important to evaluate the expected storage ring performance in user operation, while the dynamic aperture would be our concern for the lattice before correction to see if it is easily possible to establish a stored beam at an early stage of commissioning. The lifetime computation with FTT would be applicable for the corrected lattice, where the agreement is better (Fig.~\ref{fig:TLMATvsTLFTT_afterLOCO}), and attractive because of its short computation time. 

\subsection{Fully coupled lattices}\label{sec:CoupledLattice}
So-called ``round-beam modes,'' where the natural emittance is shared between the horizontal and vertical planes, is interesting for NGLS due to a longer Touschek lifetime and a lower horizontal emittance even though the photon beam brightness is degraded~\cite{xiao:ipac15-mopma013, hidaka:ipac18-tupmk018, marti:ipac22-weozsp4, kallestrup:ipac23-mopm021}. The most common approach to create a round beam is bringing the betatron working point close to or onto the linear difference resonance, $Q_x - Q_y = p$, and introducing a small amount of transverse coupling with, e.g., skew quadrupoles. The horizontal and vertical particle oscillations are then fully coupled, and thus the second approximation in FTT (that the lattice is uncoupled, see Sec.~\ref{sec:interface}) is not justified.
It can, however, be replaced by the following two conditions:
\begin{enumerate}
\item The vertical dispersion is negligible along the storage ring.
\item The magnitude of the coupling coefficient for the linear difference resonance, $|C^-|$, fulfills $|C^-| \ll 1$.
\end{enumerate}

The first condition ensures that the Touschek scattered particle has a set of initial coordinates where $y=y'=0$ is approximately true. It is generally important not to degrade the photon beam performance unnecessarily, hence, the residual vertical dispersion should be suppressed before introducing the coupling. Imposing this condition is therefore not only for the sake of FTT but also for the proper setup of round beam.

Concerning the second condition, the coupling coefficient for the linear difference resonance when the tunes are on-resonance is given by 
\begin{equation}
 	C^- = \frac{1}{2\pi}\oint k_s(s)\sqrt{\beta_x(s)\beta_y(s)}e^{-i[\phi_x(s)-\phi_y(s)]} \text{ds},
\end{equation}
where $k_s$ is the normalized skew quadrupole strength. From this coefficient, we can calculate the number of turns required for the action of the horizontal betatron motion to be transferred to the vertical plane as $|C^-|^{-1}$~\cite{guignard_1977}. Therefore the second condition ensures that, for the scattered particle tracked to the reference point, the vertical coordinates are approximately $y=y'=0$ when the first condition is simultaneously fulfilled. In practice, $|C^-| \ll 1$ is well suited for round-beam operation. When these conditions are met, the $x$--$x'$--$\dpp$ polyhedron should be valid since it is computed with initial vertical coordinates of $y=y'=0$. 

To confirm if the above replacement is valid, we examin a fully coupled lattice. The SLS 2.0 lattice is modified to move the working point onto the coupling resonance~\cite{kallestrup:ipac23-mopm021}. A set of skew quadrupoles are used to introduce a moderate coupling of $|C^-| = 0.01$: the transverse actions therefore exchange within 100 turns. These skew quadrupoles are in the nondispersive sections, and the vertical dispersion is kept zero all around the ring. The DA polyhedron is computed for a number of turns of 1024, and thus includes the effects of the coupling. The horizontal and vertical emittances of the SLS 2.0 lattice vary, at full coupling, from \SI{150}{\pico\meter\radian} and \SI{10}{\pico\meter\radian}, respectively, to \SI{96}{\pico\meter\radian} in both planes. The lifetimes are computed with/without coupling and summarized in Table~\ref{tab:ResultFullyCoupled}. No machine errors are included in this benchmarking.

\begin{table}[!ht]
	\caption{Touschek lifetimes of the modified lattice with or without coupling.}
    \centering
    \begin{ruledtabular}
    \begin{tabular}{lcS[table-format=1.2]}
    & {TL [hours]}  \\
    \toprule
    Standard tracking, $|C^-| = 0.00$& 3.80 \\
    FTT, $|C^-| = 0.00$  & 3.56 \\
    Standard tracking, $|C^-| = 0.01$  & 11.0 \\
    FTT, $|C^-| = 0.01$ & 10.6  \\
    \end{tabular}
\end{ruledtabular}
    \label{tab:ResultFullyCoupled}
\end{table}
Again, we observe a good agreement between standard MA tracking and FTT both with and without coupling. FTT consistently underestimates the lifetime by 4--6\% as was observed in Sec.~\ref{sec:FTT-AT}. This indicates to us that the polyhedron in $x$--$x'$--$\dpp$ space computed as in the uncoupled lattice is valid even in the fully coupled lattice as long as the above-listed conditions are met.

\section{Conclusion}
\label{sec:Conclusion}
We have developed new methods for efficient computation of circular accelerator lattices characteristics , with a focus on the NGLS storage ring lattice. A new algorithm for fast dynamic aperture calculations, flood-fill, was introduced. The algorithm aims to find unstable initial coordinates rather than stable ones, so as to minimize the total tracking time. The flood-fill algorithm was implemented in the OPA accelerator code and benchmarked against common DA algorithms such as grid-probing, where it was found to be faster by a factor 6 to 16, depending on the shape of the dynamic aperture. Another algorithm based on a similar principle, reverse scan, was also examined and compared to the commonly used binary search algorithm; reverse scan proved to be faster than binary search by a factor 2.5--3.3.

Furthermore, the Fast Touschek Tracking (FTT) algorithm has been developed aiming at significantly speeding up the calculation of momentum acceptance. FTT reduces the extensive element-by-element computation of momentum acceptance to an evaluation whether Touschek-scattered particles will lie within a precomputed stable $x$--$x'$--$\dpp$ volume at a reference location, which is referred to as the ``polyhedron'' throughout the paper. The MA evaluation was performed by tracking scattered particles with momentum offsets from each location of the ring to the reference point, i.e., less than one turn and therefore not time-consuming. Finding the stable volume is then the process that takes most computation time in FTT. The process is, however, essentially a series of DA calculations, where the faster algorithms that we introduce can be applied.

FTT was implemented in two accelerator codes, namely OPA and Accelerator Toolbox. Thorough benchmarking of both speed and accuracy was performed, revealing 1--2 orders of magnitude speed-up with respect to the standard momentum acceptance calculation methods. We examined the lattices with errors and a fully coupled lattice. For the latter, we narrowed down conditions that enable FTT; these conditions are normally fulfilled when the storage ring is properly set up for round-beam operation.

During the benchmarking, it was found that the accuracy of the computed lifetime can depend on the algorithm used for the DA computation, particularly when the border of the DAs used to construct the polyhedron is not completely smooth.

The largest discrepancy between standard tracking and FTT that we observed was about 25\% while it was about 7\% on average over many machines with random errors. Therefore, FTT would be applicable for the evaluation of the expected lattice performance. Further, the FTT algorithm may be well suited for numerical lattice optimization procedures that repeat DA and lifetime computations numerous times.

\section{Outlook\label{sec:Outlook}}
The presented work leads to significant improvements in the calculation of dynamic aperture and momentum acceptance. There are still several other ideas that are worth exploring. \\

\paragraph{Polar flood-fill.} The flood-fill algorithm searches for unstable points on a rectilinear grid to arrive at the required representation of a 2d DA region. However, it is possible to perform the flood fill algorithm not in $x$--$x'$ plane but directly in the $r, \theta$ space. While polar grids have nonconstant point density over the plane, further research on this problem could remove the need for polygon conversion altogether. The polar flood-fill may ease the interpolation along $\dpp$ when reconstructing the 3d DA used in the FTT algorithm.\\

\paragraph{3d flood-fill.} Flood filling the 2d DA slices can resolve shapes more accurately than binary search due to the underlying assumptions. There could be, however, extreme cases (like an ``O''-shaped DA slice) where 2d flood fill would fail. Extending the method to three dimensions is conceptually simple and could resolve even more complicated shapes, albeit at a cost of increased computational effort. With such an algorithm, the 3d DA is automatically granted.\\

\paragraph{Utilizing multiple plane crossings of each trajectory.}
During DA computations, the evaluation location plane is crossed many times, especially for captured particles. Even with flood-fill, only the coordinates of the first crossing of that reference plane are used explicitly. More information could be utilized in principle. For the interface approach to work in this setting, it would pose the additional requirement that the motion should take place fully in the reduced interface dimensions, limiting this additional extension to 4d tracking (without synchrotron motion) only.


\clearpage
\newpage
\appendix

\section{Flood-fill algorithm applied to dynamic apertures}\label{app:floodfill}
\subsection{Algorithm details}
The computational effort required to compute the DA in a given plane can be expressed via the number of total particle turns that need to be computed. There is a known asymmetry in computational effort for particles that are stable and those that are lost; by construction, lost particles quickly terminate their tracking and are thus cost effective in terms of computation.

The approach of flood-fill is to avoid tracking particles within the aperture as much as possible. To do this, we consider the 2d aperture as a pixel image and then, starting from one or two corners of the image, use the fill tool to paint the unstable region. The fill tool is advantageous, as it will stop at the boundary of the stable region and not go inside, thus potentially suppressing the computation of a large number of stable pixels.
A comparison between the known raster graphics algorithm (Alg.~\ref{alg:floodfill_image} shows a common, queue-based version of \cite{floodfill}) and the modified algorithm for apertures (Alg.~\ref{alg:floodfill_dynap}) can be performed. In the former algorithm filling raster graphics, the image is represented by the array $C$, and is both the input and the output of the algorithm. In the latter, the input is a tracking function, and the output is a monotone image $M$. As the goal of the algorithm is to prevent unnecessary tracking, the bookkeeping is done using $M$ only.

\begin{algorithm}
\caption{Flood-fill algorithm for painting an orange background area in the image $C$ red.}
\label{alg:floodfill_image}
\begin{algorithmic}
\State queue $\gets$ empty list of 2d pixel positions
\State queue $\gets$ append start pixel position
\While{queue is not empty}
  \State $(p,q)$ $\gets$ pop first element of queue
  \If{$p,q$ can index $C$ and $C_{p,q} ==$ orange}
      \State $C_{p,q} \gets$ red 
      \State queue $\gets$ append $(p+1,q)$
      \State queue $\gets$ append $(p-1,q)$
      \State queue $\gets$ append $(p,q+1)$
      \State queue $\gets$ append $(p,q-1)$
  \EndIf
\EndWhile
\end{algorithmic}
\end{algorithm}

\begin{algorithm}
\caption{Modified flood-fill algorithm for storing a 2d aperture image in a matrix $M$. The tracking function $f_N(p,q)$ returns the number of survived turns at pixel $(p,q)$, stopping at $N$ turns.}
\label{alg:floodfill_dynap}
\begin{algorithmic}
\State $M \gets -1$ for all entries \Comment{-1: uncomputed values}
\State queue $\gets$ empty list of 2d pixel positions
\State queue $\gets$ append start pixel position
\While{queue is not empty}
  \State $(p,q)$ $\gets$ pop first element of queue
  \If{$p,q$ can index $M$ \textbf{and} $M_{p,q} == -1$}
  \State $S \gets f_N(p,q)$
  \State $M_{p,q} \gets S$
  \If{$S \neq N$} \Comment{particle was lost}
      \State queue $\gets$ append $(p+1,q)$
      \State queue $\gets$ append $(p-1,q)$
      \State queue $\gets$ append $(p,q+1)$
      \State queue $\gets$ append $(p,q-1)$
  \EndIf
  \EndIf
\EndWhile
\end{algorithmic}
\end{algorithm}

\subsection{Expected time and memory complexity}
\label{sec:ccomp}

When increasing the maximum number of computed turns $N \to \infty$, the computation effort for particles that are lost becomes negligible in comparison. The pixels are only computed along the border of the aperture region and not inside. Therefore, we expect the direct computation of 2d aperture to scale quadratically with resolution in a linear direction, while (excluding fractal-like aperture boundaries) the flood-fill computation scales linear with this resolution.

\subsection{Results}
We first computed the DA for various lattices with grid probing and recorded the number of tracked turns on a 2d grid. The flood-fill algorithm was simulated using the map as input instead of repeating the tracking. To measure efficiency, we then compared the total number of turns in grid probing with the number of turns in the simulated flood-fill.

\begin{figure*}
    \centering \includegraphics[width=\textwidth]{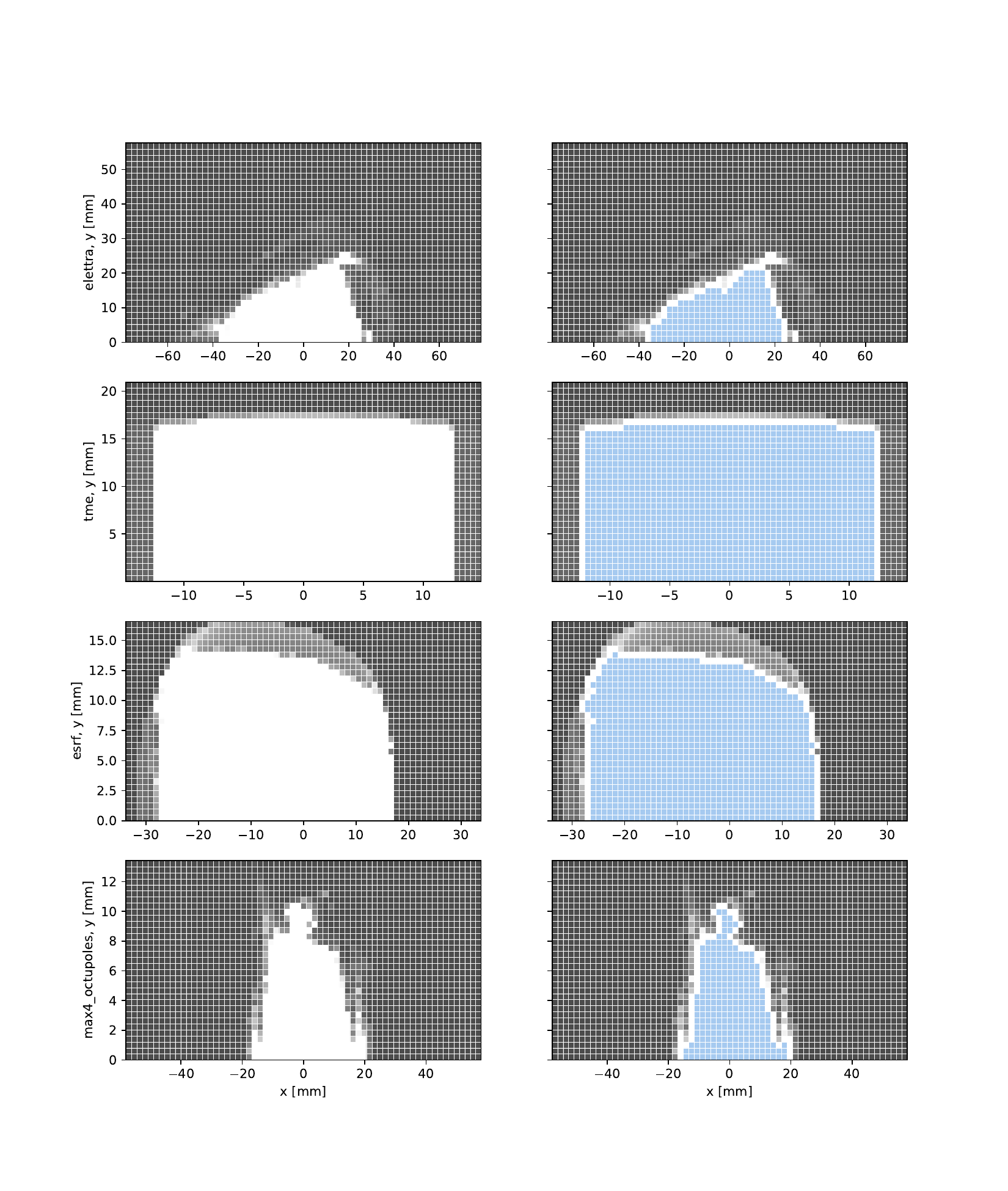}
    \caption{Raster plots for 2d DAs, the grayscales are on log2, sky blue means uncomputed. Left: Fully-computed pixel maps. Right: Result of flood-fill algorithm. Preliminary plot: shown are OPA example files. 
    The four accelerator lattices are examined, Elettra, Theoretical Minimum Emittance (TME) type lattice, ESRF and MAX IV from the top to the bottom.
    }
    \label{fig:2d_comparison}
\end{figure*}

Figure~\ref{fig:2d_comparison} shows a comparison of the maps computed by the direct procedure computing all pixels (grid probing), and the flood-fill algorithm. The required tracking effort is presented in Tab.~\ref{tab:2d_comparison}. In Fig.~\ref{fig:2d_comparison}, the region of DA search in the $x$--$y$ plane was set rather arbitrarily, and it is seen that the flood-fill algorithm is more efficient when the search region is better adjusted, i.e., the DA is not too small with respect to the search region.

\begin{table}[h]
	\caption{Required tracking effort, measured in number of computed turns for full-map vs flood-fill.}
    \centering
    \begin{ruledtabular}
    \begin{tabular}{l|ccc}
         Lattice & Full tracking & Flood fill & Ratio  \\ 
         \toprule
         Elettra & 245652 & 54874 & 0.22  \\
         TME & 1471380 & 107380 & 0.07  \\
         ESRF & 1120046 & 99046 & 0.09  \\
         MAX IV & 337859 & 75793 & 0.22  \\
    \end{tabular}
\end{ruledtabular}
    \label{tab:2d_comparison}
\end{table}

\section{Pitfalls to acceptance calculations}\label{AppendixIslands}
A common problem in the calculation of dynamic apertures, either in $x$--$y$ or $x$--$x'$ space, is when stable ``satellite'' areas outside of the ``main'' stable DA are present. These stable areas typically appear as either stable islands or as chaotic (or ``fuzzy'') regions in phase space. Different algorithms will lead to different DA estimates, as was already seen for the $\dpp=6\%$ DA in Fig.~\ref{fig:pda}. 

\begin{figure}
    \centering
    \includegraphics[width=0.45\textwidth]{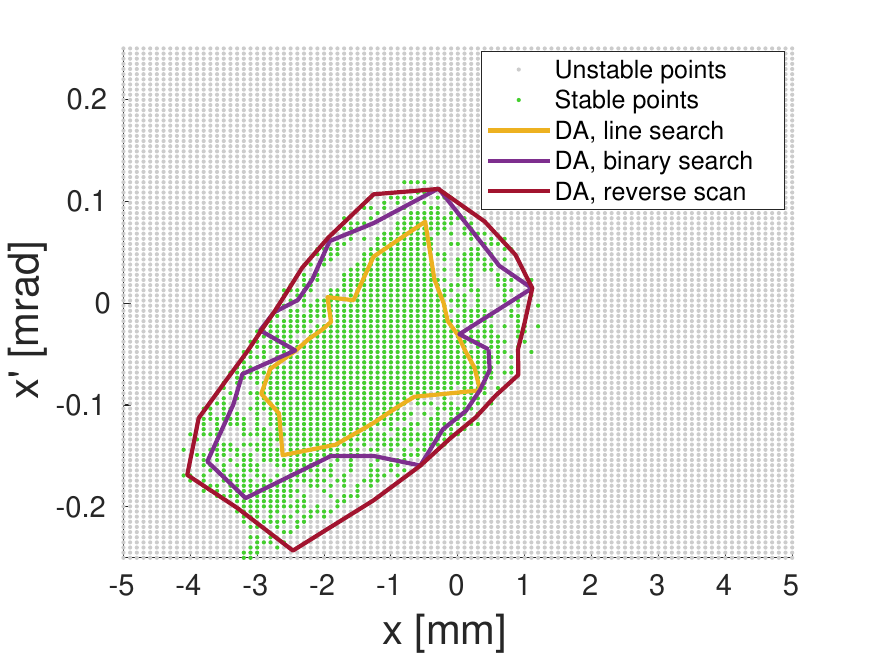}
    \caption{Stable (green) and unstable (gray) areas of $x$--$x'$ phase space for the SLS 2.0 lattice with a seed of sextupole errors leading to a ``fuzzy'' dynamic aperture. } 
    \label{fig:FuzzyDA}
\end{figure}

When these misestimated DAs are used for constructing the polyhedron in FTT, it will lead to incorrect estimates of the MA. As a prominent example, we calculate the stable areas in the $x$--$x'$ DA at $\dpp=-5.6\%$ for one of the seeds used for the computation in Fig.~\ref{fig:ATLifetimeComparison}, i.e., including strong sextupolar errors. Additionally, we apply the DA algorithms line search, binary search and reverse scan. The results are shown in Fig.~\ref{fig:FuzzyDA}. It is seen that reverse scan overestimates the DA, while line search underestimates. The binary search results in a DA between the two. From this result, we see how defining a smooth $x$--$x'$-$\dpp$ polyhedron
may not be the perfect representation of the stable initial coordinates.

The essence of the flood-fill algorithm is to mainly compute the unstable areas, where the tracking terminates quickly. The algorithm stops the search when the stable outer boundary is explored. The pitfalls are dynamic apertures where an unstable region is fully surrounded by stable areas; in this situation the algorithm will not recognize such unstable regions. A constructed, but realistic, example hereof is a phase space that has 
multiple stable fixed-points (SFPs). Two examples that utilize it are the Multi-Turn Extraction technique~\cite{mte2002,mte2017}, occasionally used in proton synchrotrons, and the transverse resonance island buckets scheme, which has recently gained significant interest 
in the synchrotron light source community~\cite{ries:ipac15-mopwa021,OLSSON2021165802,holldack_2020,holldack_2022}.

\begin{figure}
    \centering
    \subfloat{\includegraphics[width=0.4\textwidth]{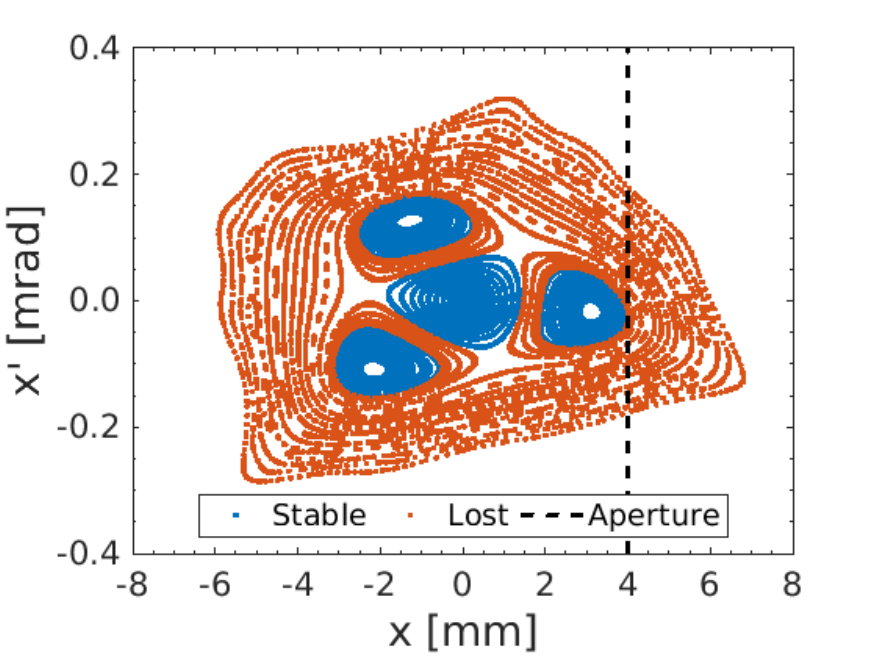}\label{fig:phaseSpace_tribs_y0mm}}
    \vspace{1pt}
    \subfloat{\includegraphics[width=0.4\textwidth]{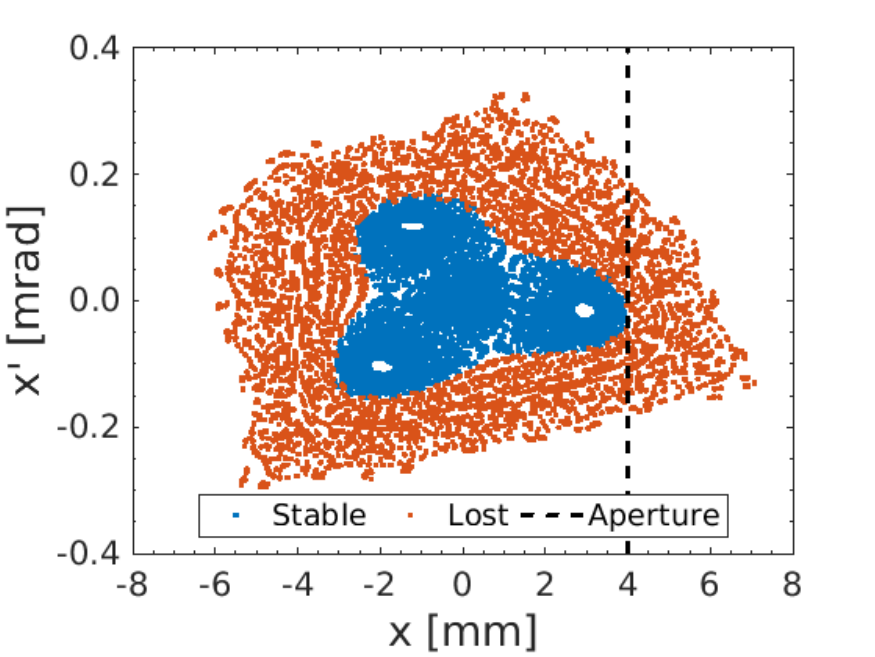}\label{fig:phaseSpace_tribs_y1mm}}
    \caption{Horizontal phase space of modified SLS 2.0 lattice close to the $3\nu_x = n$ resonance to create stable Transverse Resonance Island Buckets. The dashed line indicates a hypothetical physical aperture to create unstable areas of dynamic aperture. top: $y=\SI{0}{\milli\meter}$. Bottom: $y = \SI{1}{\milli\meter}$.} 
    \label{fig:phaseSpace_tribs}
\end{figure}

\begin{figure}
    \centering
    \subfloat[Grid-probing]{\includegraphics[width=0.4\textwidth]{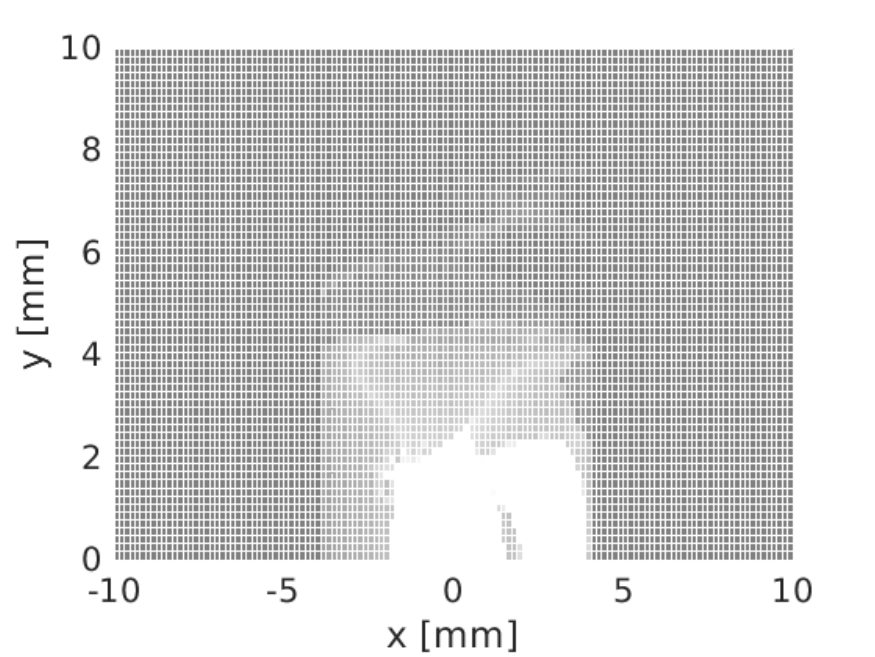} }
    \vspace{1pt}
    \subfloat[Flood-fill]{\includegraphics[width=0.4\textwidth]{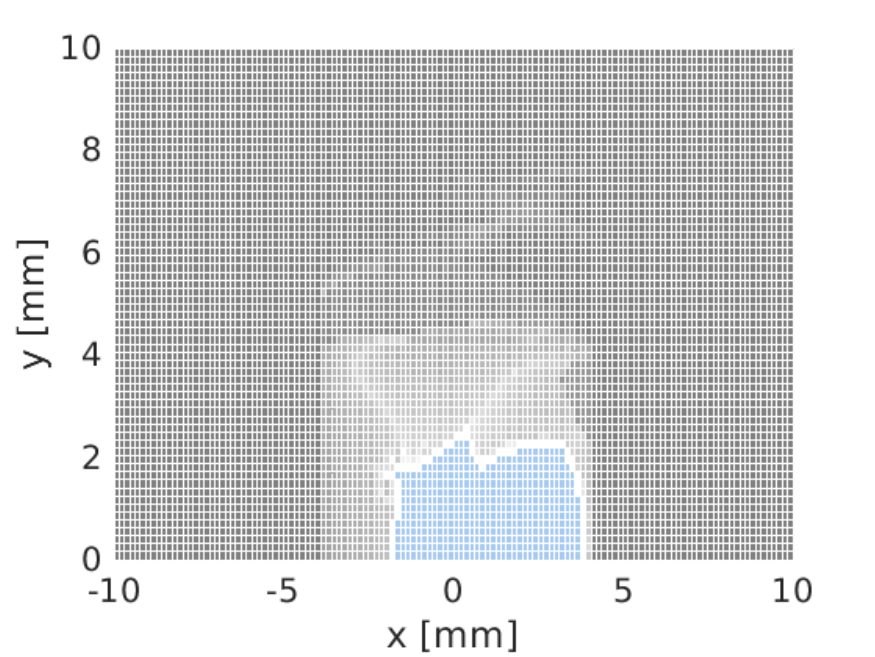}}
    \caption{Acceptances for SLS 2.0 TRIBs example. (a) Grid-probing algorithm. (b) Flood-fill algorithm.} 
    \label{fig:dynap_tribs}
\end{figure}

For illustration, we have simulated a machine based on the SLS 2.0 lattice. The working point is set close to the $3Q_x$ resonance and the sextupolar symmetry has been broken to excite the resonance: three SFPs are created in phase space. A simple aperture is then inserted at $x = +\SI{4.0}{\milli\meter}$. Figure~\ref{fig:phaseSpace_tribs} shows horizontal phase space plots from tracking simulation for two cases: one with no vertical motion and the other with a vertical motion given by an initial vertical offset of $y = \SI{1}{\milli\meter}$. As seen in Fig.~\ref{fig:phaseSpace_tribs} (a), the physical aperture cuts the acceptance into four pieces; this does not happen without the SFPs arising from the resonance. The machine has non-zero cross-term ADTS, $\frac{\partial Q_x}{\partial J_y}$ with $J_y$ being the vertical action. The positions of the SFPs are slightly shifted with the vertical motion as in Fig.~\ref{fig:phaseSpace_tribs_y1mm} and, in this case, the acceptance is cut into pieces (although the separatrix is still there). This combination of the dynamics and the physical aperture results in an acceptance with a hole in the $x$--$y$ plane as shown in Fig.\ref{fig:dynap_tribs}. GP is capable of finding the inner unstable area in $x$--$y$ plane, while FF fails. 

\begin{figure}
    \centering
    \subfloat[Uncoupled]{\includegraphics[width=0.4\textwidth]{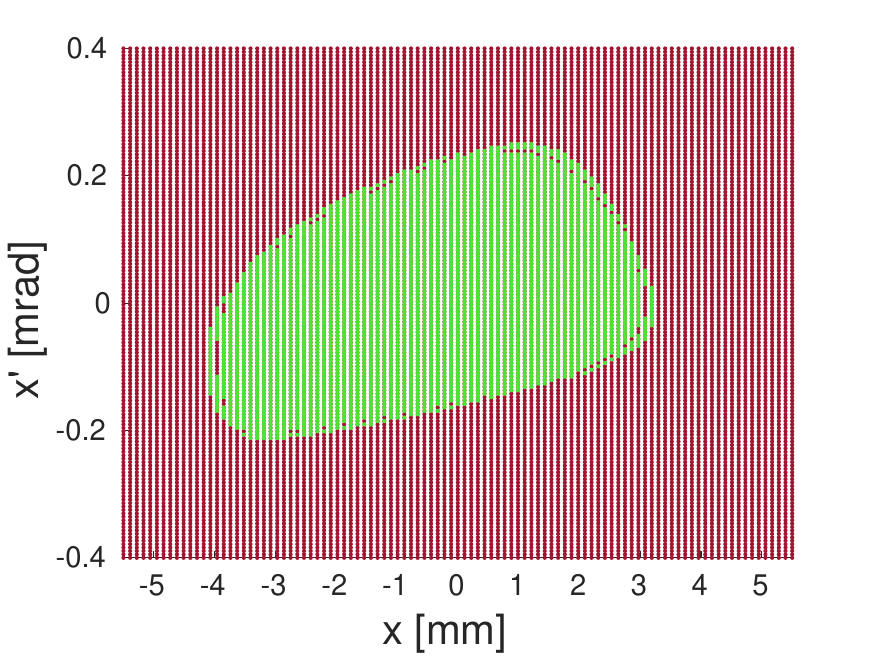} }
    \vspace{1pt}
    \subfloat[Coupled]{\includegraphics[width=0.4\textwidth]{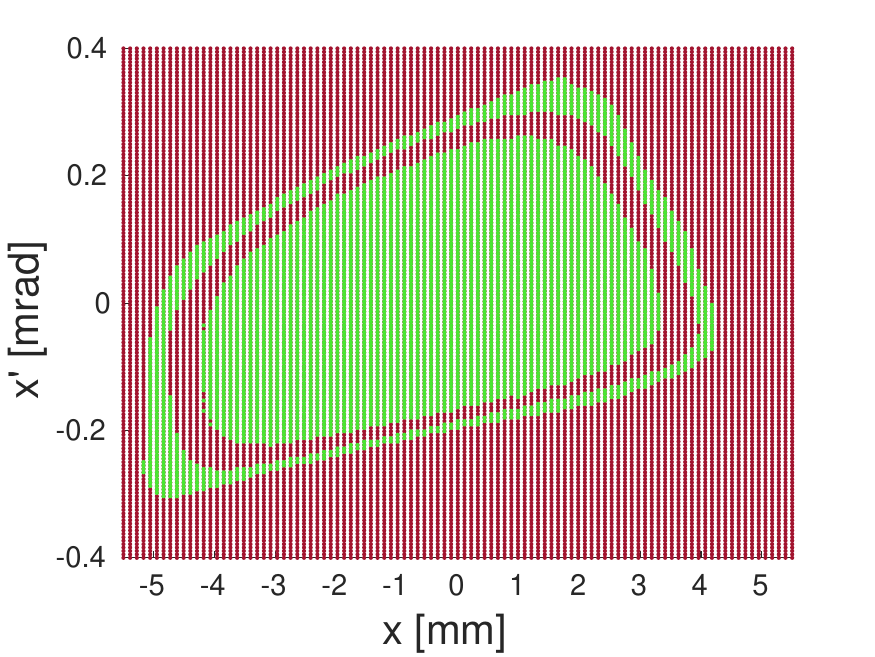}}
    \caption{Stable regions of $x$--$x'$ for the SLS 2.0 round beam lattice for $\dpp=-5.6\%$ a) without coupling and b) with coupling. Green and red dots indicate stable and unstable coordinates, respectively.} 
    \label{fig:unstableIslands}
\end{figure}

Another similar example has been observed in the SLS 2.0 round beam lattice Sec.~\ref{sec:CoupledLattice}) for $\dpp=-3.4\%$. Figure~\ref{fig:unstableIslands} shows the stable and unstable points in the $x$--$x'$ phase space with and without coupling. The particles are unstable for $x>\SI{+3.3}{\milli\meter}$ (and $x'=0$), but the introduced coupling helps stabilizing the particle motion for a small region of $\SI{4.0}{\milli\meter}<x<\SI{4.2}{\milli\meter}$, thereby giving an outer ``stable band.'' Flood fill will not be able to find the inner unstable band, nor any of the other algorithms we examined (line, reverse or binary search) except for GP.

\section{Algorithms for standard momentum acceptance tracking}
\subsection{Line search}\label{app:linesearchMA}
The line-search algorithm is the simplest, but also most time consuming approach to calculate the momentum acceptance at some location in the lattice. At a given location, a particle is started on the closed orbit with some small $\dpp$ offset. The value of $\dpp$ is incremented in small steps $\delta$, equal to the requested resolution of the momentum acceptance search, until an unstable $\dpp$-value is encountered.
The last stable $\dpp$-value is recorded as the momentum acceptance. Pseudocode for the algorithm is given in Alg.~\ref{alg:linesearchMA}.

The line-search algorithm is not prone to getting stuck in stable islands, since it checks all values of $\dpp$ with $\delta$ spacing. This makes the algorithm appear more favorable, in a sense of avoiding overestimations, to the binary search in cases where islands may appear. 

\begin{algorithm}
\caption{Line-search algorithm for momentum acceptance estimation}
\label{alg:linesearchMA}
\begin{algorithmic}
\For{every relevant position $s_p$}
\State $\vec{X}_0$ $\gets$ particle coordinates of closed orbit 
\State $\dpp$ $\gets$ 0 \Comment{momentum offset to be checked}
\State \texttt{particle\_stable} $\gets$ true \Comment{boolean indicating stability}
\While{\texttt{particle\_stable} AND $\dpp \leq \dpp_\text{max}$}
    \State  $\dpp$ $\gets$ $\dpp+\delta$ 
    \State $\vec{X}$ $\gets$ $\vec{X}_0 +  \dpp$
    \State \texttt{particle\_stable} $\gets$ \textbf{tracking subroutine}($\vec X$)
\EndWhile
\State MA$(s_p)$ $\gets$ $\dpp-\delta$
\EndFor
\end{algorithmic}
\end{algorithm}

\subsection{Sieve search}\label{app:sieve-search}
It is possible to speed up the line-search algorithm using a two-step approach, Coarse and fine scans, resembling the process of sieving with multiple mesh sizes: first, the line search is applied with a large value of $\delta=\delta_\text{C}$. This provides a momentum acceptance estimate of MA$(s_p)_\text{C}$. Next, a line search is applied for the range $\text{MA}(s_p)_\text{C}-\delta_\text{C} \leq \delta \leq \text{MA}(s_p)_\text{C}+\delta_\text{C}$ with small steps of $\delta = \delta_\text{F}$. The final results is a momentum acceptance estimate with resolution $\delta_\text{F}$. If $\delta_\text{C}$ is larger than the stable $\dpp$-region of any islands then the sieve search will lead to exactly the same results as line search. 

\bibliographystyle{unsrt}
\bibliography{main.bib}

\end{document}